\documentclass[twocolumn]{aastex63}
\pdfoutput=1 %for arXiv submission
\usepackage{amsmath,amstext,amssymb,mathtools,graphicx,color}

\usepackage{graphicx}	% Including figure files
\usepackage{amsmath}	% Advanced maths commands
\usepackage{amssymb}	% Extra maths symbols
\usepackage{xcolor}
\usepackage{lipsum}
\usepackage[normalem]{ulem}

\received{June 1, 2019}
\revised{January 10, 2019}
\accepted{\today}
\submitjournal{ApJ}

\shorttitle{High z GRBs}
\shortauthors{Fryer et al.}

\graphicspath{{./}{figures/}}
\begin{document}

\title{Properties of High-Redshift GRBs}

\correspondingauthor{Chris L. Fryer}
\email{fryer@lanl.gov}

\author[0000-0003-2624-0056]{Chris L. Fryer}
\affiliation{Center for Theoretical Astrophysics, Los Alamos National Laboratory, Los Alamos, NM, 87545, USA}
\affiliation{Computer, Computational, and Statistical Sciences Division, Los Alamos National Laboratory, Los Alamos, NM, 87545, USA}
\affiliation{The University of Arizona, Tucson, AZ 85721, USA}
\affiliation{Department of Physics and Astronomy, The University of New Mexico, Albuquerque, NM 87131, USA}
\affiliation{The George Washington University, Washington, DC 20052, USA}

\author[0000-0002-7851-9756]{Amy Y. Lien}
\affiliation{University of Tampa, Department of Chemistry, Biochemistry, and Physics, 401 W. Kennedy Blvd, Tampa, FL 33606, USA}
\affiliation{Center for Research and Exploration in Space Science and Technology (CRESST) and NASA Goddard Space Flight Center, Greenbelt, MD 20771, USA}
\affiliation{Department of Physics, University of Maryland, Baltimore County, 1000 Hilltop Circle, Baltimore, MD 21250, USA}

\author[0000-0000-0000-0000]{Andrew Fruchter}
\affiliation{Space Telescope Science Institute, 3700 San Martin Drive Baltimore, MD 21218, USA}

\author[0000-0001-5876-9259]{Giancarlo Ghirlanda}
\affiliation{INAF – Osservatorio Astronomico di Brera, Via E. Bianchi 46, I-23807 Merate, Italy}
\affiliation{INFN – Sezione di Milano-Bicocca, Piazza della Scienza 3, I-20126 Milano, Italy}

\author[0000-0002-8028-0991]{Dieter Hartmann}
\affiliation{Department of Physics and Astronomy, Clemson University, Clemson, SC 29634, USA}

\author[0000-0002-9393-8078]{Ruben Salvaterra}
\affiliation{INAF-Istituto di Astrofisica Spaziale e Fisica cosmica, via Alfonso Corti 12, 20133, Milano, Italy}

\author[0000-0000-0000-0000]{Phoebe R. Upton Sanderbeck}
\affiliation{Center for Theoretical Astrophysics, Los Alamos National Laboratory, Los Alamos, NM, 87545, USA}
\affiliation{Applied Physics Division, Los Alamos National Laboratory, Los Alamos, NM, 87545, USA}

\author[0000-0000-0000-0000]{Jarrett L. Johnson}
\affiliation{Center for Theoretical Astrophysics, Los Alamos National Laboratory, Los Alamos, NM, 87545, USA}
\affiliation{Applied Physics Division, Los Alamos National Laboratory, Los Alamos, NM, 87545, USA}



\begin{abstract}

The immense power of gamma-ray bursts (GRBs) make them ideal probes of the early universe.  By using absorption lines in the afterglows of high-redshift GRBs, astronomers can study the evolution of metals in the early universe.  With an understanding of the nature of GRB progenitors, the rate and properties of GRBs observed at high redshift can probe the star formation history and the initial mass function of stars at high redshift.  This paper presents a detailed study of the metallicity- and mass-dependence of the properties of long-duration GRBs under the black-hole accretion disk paradigm to predict the evolution of these properties with redshift.  These models are calibrated on the current GRB observations and then used to make predictions for new observations and new missions (e.g. the proposed Gamow mission) studying high-redshift GRBs.

\end{abstract}


\section{Introduction} 
\label{sec:intro}

Understanding the star formation history and the nature of stars formed at high redshift is critical to understanding galactic chemical evolution as well as the formation and evolution of galaxies, the drivers behind re-ionization and even the formation of supermassive black holes.  Although a number of probes of star formation exist, stellar explosions are one of the most powerful.  The explosions produced by these stars are sufficiently luminous to be observed at high distances, making them ideal probes of the early universe.  The most well-known example of such a probe is the use of type Ia supernovae to measure the cosmological constant~\citep{1998AJ....116.1009R, 1999ApJ...517..565P}.  However, the explosions from the collapse of massive stars may prove to be even more powerful probes at the higher redshifts.  Stars more massive than $\sim8-10$\,M$_\odot$ produce type Ib/c and type II supernovae, pair-instability supernovae and long-duration gamma-ray bursts.  Both collapse supernovae~\citep{2002ApJ...566L..63H} and gamma-ray bursts (GRBs)~\citep{1997A&A...328L..21A, Ghirlanda2004} 
have been studied as standard candles.  But they are much more powerful than that.  The short, and perhaps more importantly, the well-characterized life-times of these massive stars means that we can directly connect their explosions to the star formation history.  GRBs are particularly powerful because their progenitors are even more massive than supernovae, so they serve as both probes of the star formation history and the initial mass function of these stars~\citep{2000ApJ...536....1L,2002ApJ...574..554L}.  

However, using astrophysical explosions to probe the early universe has its difficulties.  These transients are produced by massive stars and, as such, trace the star formation of massive stars, not star formation as a whole.  In addition, we do not know the exact progenitors of these explosions.  Understanding the nature of the progenitor behind type Ia supernovae remains an active area of research~\citep{2018PhR...736....1L}.  Normal core-collapse supernovae are produced from massive stars that undergo core-collapse, setting a lower limit on supernova-producing massive stars.  This lower limit corresponds to a zero-age main sequence star mass lying somewhere between 7 and 9 M$_\odot$~\citep[for a review, see][]{2003ApJ...591..288H}.  Because of the steep initial mass function for massive stars, this uncertainty in the lower limit can drastically alter the median mass of supernova progenitors.  The upper limit is dictated by whether the collapsed core can drive an explosion.  The limit is close to 20\,M$_\odot$, but depends on both mass loss and the stellar evolution~\citep{1999ApJ...522..413F,2012ApJ...749...91F,2011ApJ...730...70O}.

It can be even more difficult to tie GRBs to specific massive stars.  The black-hole accretion disk (BHAD) paradigm~\citep{1999ApJ...518..356P,1999ApJ...526..152F} for GRBs has both fit existing data and made observational predictions that have been subsequently confirmed~\citep[for a review, see][]{2019EPJA...55..132F}.  In this paper, we focus on Type I Collapsars, those formed from the so-called direct-collapse scenario of black hole formation~\citep{2003ApJ...591..288H} where the black hole forms without launching a supernova explosion.  Even if we assume these collapsars dominate the production of normal GRBs, the exact mass limits for this direct-collapse scenario are, as yet, unknown.  It is also not yet known what specific evolutionary scenarios (e.g. binary interactions) produce the high spins needed to produce GRBs~\citep{2007PASP..119.1211F}, introducing an additional uncertainty.  Finally, engine paradigms beyond the standard BHAD paradigm may also exist.  If so, such models may require an entirely different progenitor scenario.  

Despite these progenitor uncertainties, astrophysical transients remain a powerful probe of the early universe.  In this paper, we devise a self-consistent model including progenitors of a wide set of astrophysical transients, namely supernovae, hypernovae, pair-instability supernovae, and GRBs (under the BHAD paradigm).  In particular, we find that astrophysical transients are powerful probes of the redshift evolution of the initial mass function.  In section~\ref{sec:progenitor}, we describe the redshift evolution (driven by metallicity evolution) of our progenitors discussing both changes in the star formation rate and initial mass distribution of stars and the subsequent evolution of those stars.  We further study the effect these features have on the observable properties of the transients in section~\ref{sec:obs}.  Section~\ref{sec:calibration} describes our tests of our model against the observations of GRBs at low-redshift followed by a series of observational predictions for our astrophysical transients at high redshift in section~\ref{sec:highz}.  We conclude with a summary of observations that will constrain the redshift evolution of star formation.

\section{Redshift Evolution}
\label{sec:progenitor}

The properties and rates of gamma-ray bursts primarily vary with redshift both because the star formation rate and metallicity varies with redshift~\citep[for a review, see][]{2014ARAA..52..415M}.  The metallicity, in turn, alters both the initial mass function (IMF) of newly formed stars and the subsequent stellar evolution.  In this section, we review redshift effects on the star formation rate, the IMF and stellar evolution.  For each, we discuss the uncertainties in this evolution and outline the prescriptions (along with their justification) used in this study.

\subsection{Star Formation Rate} 
\label{sec:SFR}

The star formation history, in particular at high redshift, is difficult to determine.  Studies have been done using a wide range of probes (emission from star formation regions, supernovae, gamma-ray bursts) over a broad range of photon energies from the infra-red to gamma-rays.  In general, observations of the emission from star forming regions~\citep{1996MNRAS.283.1388M, 2006ApJ...651..142H,2020ApJ...902..112B} predict a much sharper drop off at high redshifts than astrophysical transients~\citep{2004MNRAS.347..942G,2015arXiv150308147W}.  From dust obscuration to the evolution of the initial mass function, uncertainties in model assumptions make it difficult to ascertain a firm star formation history at high redshift.  

For our models, we adopt the star-formation rate from \cite{2014ARAA..52..415M} as the benchmark star-formation rate in our simulation. This star-formation rate is described by the following equation,
\begin{equation}
\label{eq:sfr}
\rho = 0.01 \frac{(1+z)^{2.7}}{1+[(1+z)/2.9]^{5.6}},
\end{equation}
where $\rho$ is in unit of $M_\odot \ \rm yr^{-1} \ Mpc^{-3}$.

In addition, we also run the simulations of the main theoretical model using two other star-formation rates from
\cite{Madau17}
and
\cite{2006ApJ...651..142H}, in order to estimate how the predicted GRB rate might change due to different assumptions of the star-formation rate.  The star-formation rate from \cite{Madau17} is described as
\begin{equation}
\label{eq:sfrmad}
\rho = 0.01 \frac{(1+z)^{2.6}}{1+[(1+z)/3.2]^{6.2}},
\end{equation}
and the star-formation rate from \cite{2006ApJ...651..142H} is 
\begin{equation}
\rm{log}(\rho) = 
\begin{cases}
3.28\rm{log}(1+z) - 1.82, \ \ \ \ \ {z \leq 1.04} \\
-0.26\rm{log}(1+z) - 0.724,  \ \ {1.04 <  z \leq 4.48} \\
-f_{\rm highz}\rm{log}(1+z) + 4.99, \ \ \ \   {z > 4.48}
\end{cases}
\end{equation}

$f_{\rm highz}$ dictates the drop in star formation at high redshift.  Most observations of star formation predict that $f_{\rm highz}$ lies between $5-8$ whereas astrophysical transients predict $f_{\rm highz}$ values in the $0-4$ range.  These differences could be explained by observational biases in the different samples.  They may also be explained by an evolution in the progenitors of these transients as a function of redshift.  In the next two subsections, we study the effects of metallicity (and hence redshift) on GRB and supernova progenitors.  Although these models produce very similar results at low redshift (with just different shapes to their fits to the data), the different dependencies on the redshift at high values of the redshift can lead to very different rate predictions.

\subsection{Metallicity Effects on Star Formation}
\label{sec:IMF}

Long GRBs are produced in the most massive stars and, as such, the rate of GRBs depends sensitively on the poorly-understood initial mass distributions of massive stars, a.k.a. the initial mass function (IMF).  A number of stellar population studies in the local group of galaxies have been used to infer observed IMFs~\citep{1955ApJ...121..161S,1979ApJS...41..513M,2001MNRAS.322..231K,2003PASP..115..763C}.  These studies typically focus on stars formed at high metallicities, but the general consensus is that, as long as there are some metals to drive cooling, the IMF will not change dramatically with metallicity.  All of these studies obtain similar results for the mass distribution of stars above 1\,M$_\odot$:
\begin{equation}
f(M) \propto M^{-\alpha} dM
\end{equation}
where $f(M)$ is the fraction of stars produced within the range of $M$ and $M+dM$ and $\alpha$ is typically set to $\sim 2.3$ (e.g. for Salpeter predicted $\alpha=2.35$).  For our study, we use the Kroupa distribution to include the flattening of the IMF at low masses:
\begin{equation}
\alpha_{\rm IMF}=
\begin{cases}
    0.3,& \text{if } M<0.08 \\
    1.3,& \text{if } 0.08<M<0.5 \\ 
    2.35,&  \text{if } M>0.5
\end{cases}
\end{equation}
where stellar masses are measured in solar mass units and the IMF is normalized to unity over the stellar mass range:  0.0001 - 500\,M$_\odot$. The mean stellar mass is $\langle M \rangle \sim$ 0.7 and the fraction of  single stars that create a neutron star or black hole is F$_{cr} = 0.007$ if one assumes that all stars with a mass greater than 8.5\,M$_\odot$ experience core collapse.  This corresponds to a star formation rate of 1.5 stars per solar mass of star formation and, in the Milky Way, where the star formation rate is roughly 1.9\,M$_\odot \, {\rm yr}^{-1}$ \citep{2012ARAA..50..531K}, a SN rate of roughly 1.4 SN per century.  

Metallicity can alter the nature of star formation in a number of ways.  For example, as the metallicity decreases, metals can not efficiently cool proto-stellar clumps and the critical mass needed for clump collapse that initiates star formation increases~\citep[for a review, see][]{2013RPPh...76k2901B,2020AJ....160...78R}.  In general, it is believed that, until the metallicity is very low, metal-driven cooling is sufficiently efficient to minimize the effect on the initial mass function.  However, more recent studies argue that there is significant flattening of the IMF at a metallicity between $0.01-0.1 z_\odot$\citep{2021arXiv210304997C} corresponding to a redshift of 2.5-5 (see Section~\ref{sec:zvz}).  

In addition, metals can alter the fragmentation of stars where less fragmentation occurs (fewer binaries are produced) at lower metallicities~\cite{2001MNRAS.328..969B}.  The former effect serves to produce more massive stars, flattening the IMF and producing more GRB progenitors.  Less fragmentation also flattens the IMF, but it produces fewer binaries.  But if binary interactions are essential to the formation of GRBs, the lack of fragmentation at high redshift could reduce the overall GRB rate at low metallicities (and high redshifts).  Our understanding of star formation is far from complete.  Magnetic fields can play an important role in star formation but the effect of magnetic fields on star formation and, in particular, with redshift evolution, remains an active area of research~\citep{2020AJ....160...78R}.   At high metallicities, stellar winds are much stronger and can also alter star formation~\citep{2020MNRAS.493..447C}.

For our models, we will focus on two effects:  the flattening of the IMF at low metallicity (from the larger protestellar collapse criterion and less fragmentation) and a smaller binary fraction (from the reduced fragmentation).  For the IMF evolution, we must assume both the dependence of this flattening on the metallicity and an evolution of the metallicity with redshift (which depends on the mixing of metals).

Because we don't exactly know the evolution of the IMF with metallicity, we combine both the star formation and metallicity evolution into a single set of uncertain quantities (we do need a stand-alone metallicity/redshift prescription for our stellar evolution models, see section~\ref{sec:progenitor}).  To vary a wide range of evolutionary parameters, we use a simple paramaterized prescription for the IMF slope ($\alpha_{\rm IMF}$) for massive stars with masses above 0.5\,M$_\odot$ above a critical redshift $z_0$:
\begin{equation}
\alpha_{\rm IMF} = 2.35 - f_{\rm IMF} (z-z_0)
\label{eq:imf}
\end{equation}
where we vary $f_{\rm IMF}$ from 0-0.4 and vary $z_0$ from 2.5 to 9.  A $z_0=2.5-5$ assumes a flattening of the IMF much earlier than expected in simulations but in line with more recent results~\citep{2021arXiv210304997C}.  As we shall see in our comparison to GRB observations, either this is required or a drastic change in the currently-predicted star history at high redshift (or both) is necessary. 

\begin{figure}
\plotone{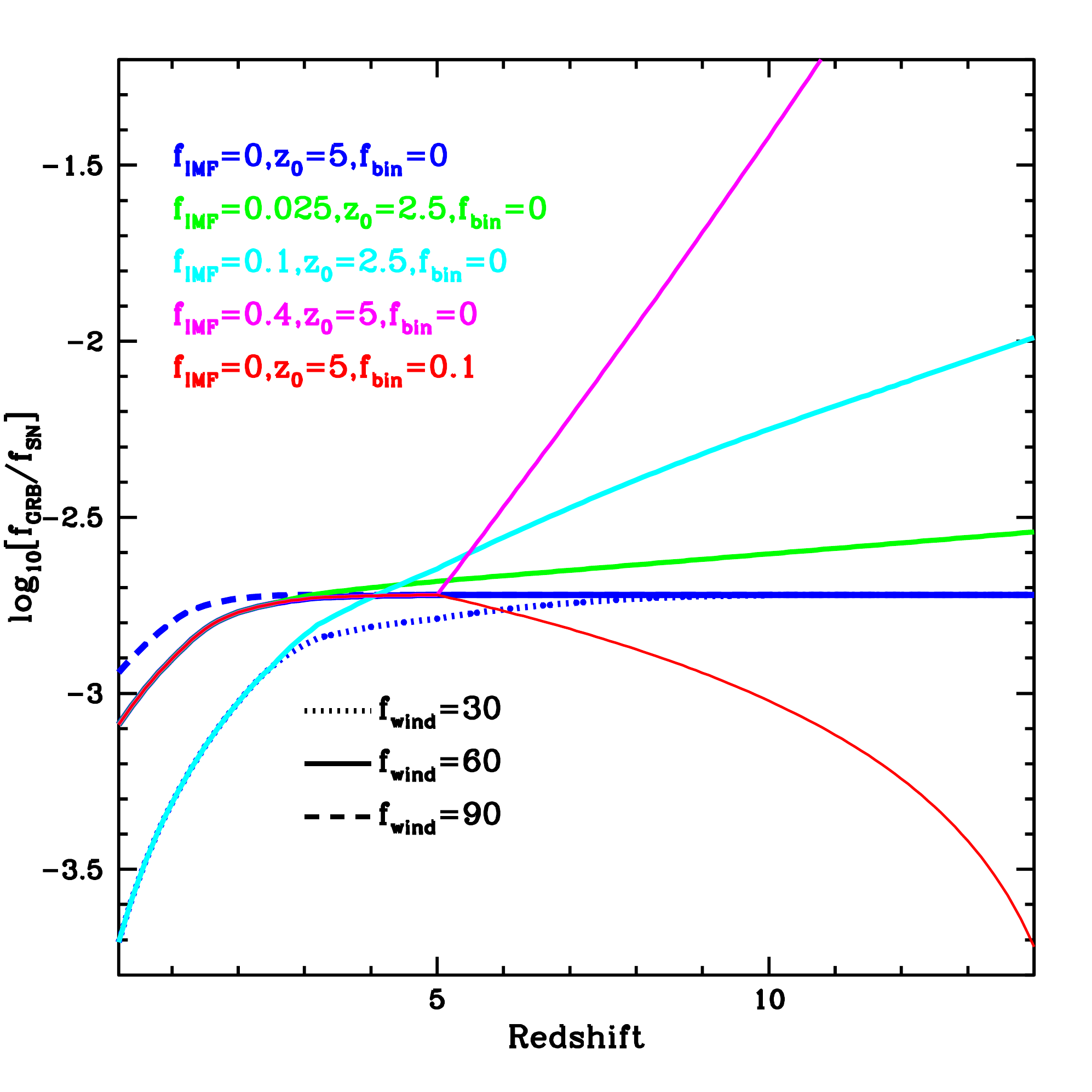}
\caption{Normal and dim GRB rate as a function of redshift for a range of values for $f_\alpha$, $f_{\rm bin}$ and $z_0$.  This plot focuses on extreme results, where we begin to alter the IMF and binary fraction at redshifts as low as 2.5-5.0.  These produce very large extremes in the total GRB to SN ratio.  At redshift of 10, the GRB rate ranges from 0.095\% to 4\% of the SN rate.  We also vary a wind parameter ($f_{\rm wind}$) to study the extreme effects of mass loss on the rate.  These models assume that only one in a hundred black hole systems have sufficient angular momentum to form a collapsar disk.  The relative GRB to SN rate can be scaled by changing this assumption, see section~\ref{sec:calibration}.  For standard wind scenarios ($f_{\rm wind} = 30-60$), the effects of mass-loss from metal-driven winds are minimal above a redshift of 2.} % caption
\label{fig:IMF}
\end{figure}

Many of the progenitor scenarios for long-duration gamma-ray bursts require binary interactions~\citep[for reviews, see][]{1999ApJ...526..152F,2007PASP..119.1211F,2019EPJA...55..132F}.  If less fragmentation occurs at low metallicities, one might expect fewer binaries and, hence, fewer GRBs~\citep{2010ApJ...708..117B}.  In addition, with less expansion in the giant phase for massive stars, even if binaries form, we expect fewer binary interactions.  We implement an additional decrement to the GRB rate with redshift:
\begin{equation}
    {\rm Fraction_{Binaries}}=1.0-f_{\rm bin} (z-z_0)
\end{equation}
where we vary $f_{\rm bin}$ from 0 to 0.1.  This interacting binary fraction will be multiplied to our long-duration GRB rate.  Figure~\ref{fig:IMF} shows the ratio of long GRBs to the supernova rate for a range of values of $f_{\rm IMF}$, $f_{\rm bin}$  and $z_0$.  Uncertainties in the IMF can produce a wide range of results at high redshift.

\subsection{Metallicity Effects on Stellar Evolution}
\label{sec:prog}

\subsubsection{Metallicity and Wind Mass Loss}

The primary effect of the metallicity on stellar evolution is its role in mass loss.  Standard mass loss in massive stars is produced through line-driven mechanisms~\citep{2000AA...360..227N,2002ApJ...577..389K}.  At solar metallicities, this mass-loss can be quite extreme, drastically altering the mass and core-structure of massive stars.  These mass-loss-metallicity effects have long been incorporated into studies of massive-star progenitors of supernovae and gamma-ray bursts~\citep{2003ApJ...591..288H,2007ApJ...670..584Y,2007PASP..119.1211F}.  These studies found that for the black hole accretion disk engine for GRBs, mass-loss prevents the formation of GRBs at high metallicity, but once the metallicity drops below $\sim 0.1$ solar metallicity, mass-loss and metallicity effects are, relative to the extreme effects at high metallicity, minor.  There is a maximum mass at which the type I Collapsar forms.  Without mass loss, above about 130\,$M_\odot$, pair-instability disrupt the star.  However, if mass loss is extensive, this limit can be set by mass loss.  As the metallicity in the star increases, stars lose too much mass to collapse to form black holes~\citep{2003ApJ...591..288H}.  Mass-loss is most effective for the most massive stars and, with increasing metallicity, mass-loss will first affect the most-massive stars, altering their final outcome from a collapsar to a normal supernova.  With increasing metallicity, the mass of stars where mass-loss alters the final outcome begins to drop.  To incorporate this effect, we assume the maximum mass of GRB formation based on the mass loss:
\begin{equation}
    M_{\rm max}^{\rm GRB} = min(130\,M_\odot,-f_{\rm wind} log_{10} (Z_{\rm metal}/z_\odot)M_\odot)
    \label{eq:wind}
\end{equation}
where $Z_{\rm metal}$ is the metallicity, $z_\odot$ is solar metallicity, and $f_{\rm wind}$ dictates the dependence of the mass loss on the metallicity.  We vary $f_{\rm wind}$ from 30-9000 where 9000 limits mass loss to near solar-metallicity stars.  For lower values, the effects of mass on. the fate of massive stars can be more extended.  We use 130\,M$_\odot$ for the absolute maximum mass of GRB formation assuming that stars undergo pair-instability above this limit.  It is worth noting that mass loss can allow these massive stars to avoid this pair-production instability.  For the IMF at high metallicities, these stars do not contribute significantly to the GRB rate and our removal of these will not affect our rate.   Beyond the pair-instability mass gap, the most massive stars (zero-age main sequence star mass above $\sim 300 M_\odot$) can form black holes.  We will discuss these Pop III collapsars below (Section~\ref{sec:lgrb})

\begin{figure}
\plotone{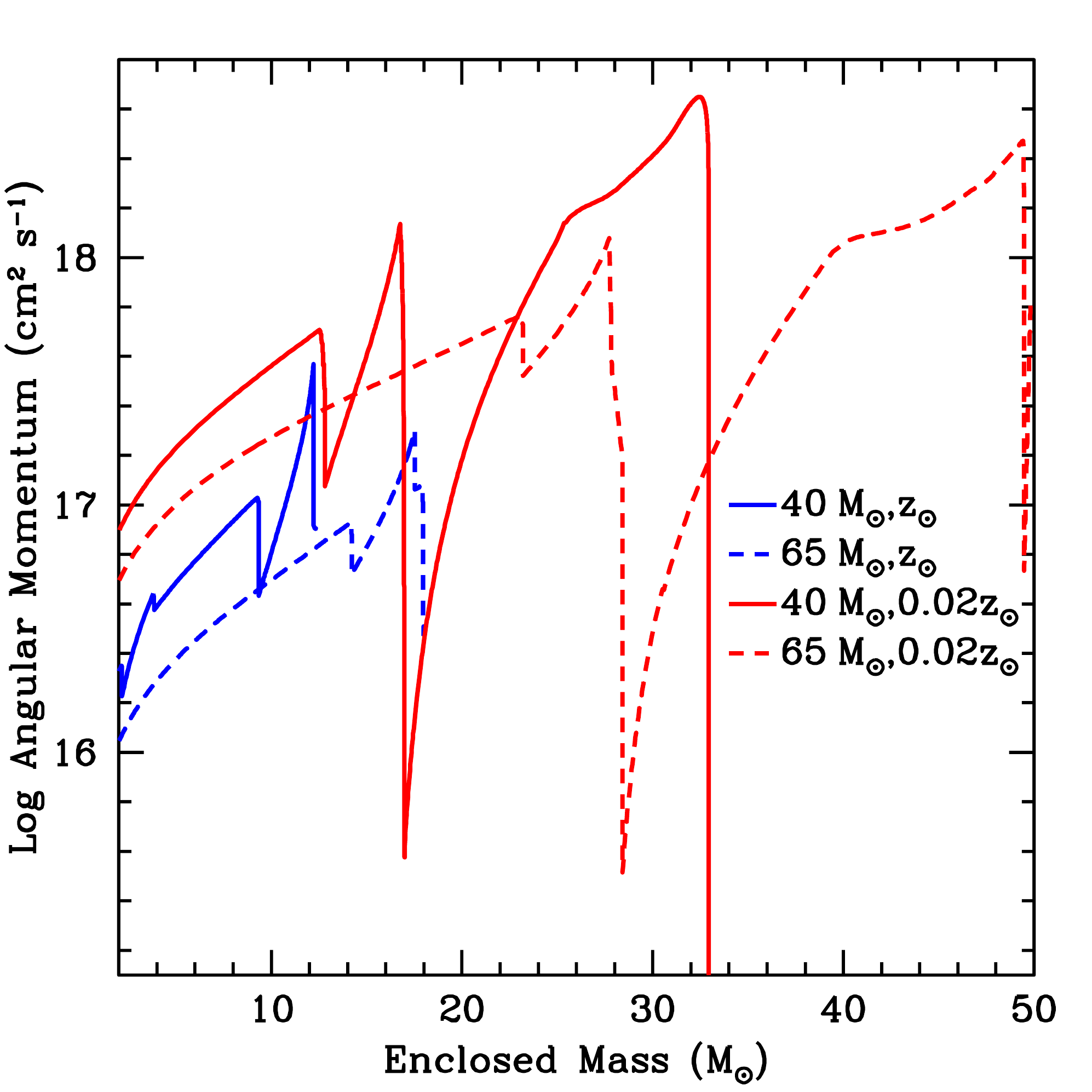}
\caption{Log of the specific angular momentum as a function of mass coordinate for a 40 and 65\,M$_\odot$ star (blue and red curves respectively)  at solar and 0.02 solar metallicities (solid, dashed curves respectively).  If this difference in angular momentum persists, we should see a change in the rate of GRBs at these modest metallicities.} % caption
\label{fig:angm}
\end{figure}

Mass loss from winds also removes angular momentum from stars.  For the fastest rotating stars, winds are efficient at carrying away angular momentum and stars with high mass loss will spin slower than those without this mass loss.  Figure~\ref{fig:angm} shows the angular momentum profiles of two massive stars (40 and 65\,M$_\odot$) at two metallicites (solar and 0.02 solar) from the GENEC code.  These calculations assume no strong coupling between layers from a magnetic dynamo, producing the high-spin progenitors needed to produce black hole accretion disks.  Even without this coupling, mass loss removes enough angular momentum to make much slower progenitors at high metallicities.  The angular momentum at lower metallicites is a factor of 4 higher than the models at solar metallicity.  If these models are correct, the GRB rate could increase as the metallicity drops from solar to 0.02 solar.  However, we note that there is growing evidence that there is some coupling between burning layers and this mass loss affect on angular momentum could be muted~\citep{2020AA...636A.104B}. 

\subsubsection{Other Metallicity Effects on Stellar Evolution}

\begin{figure}
\plotone{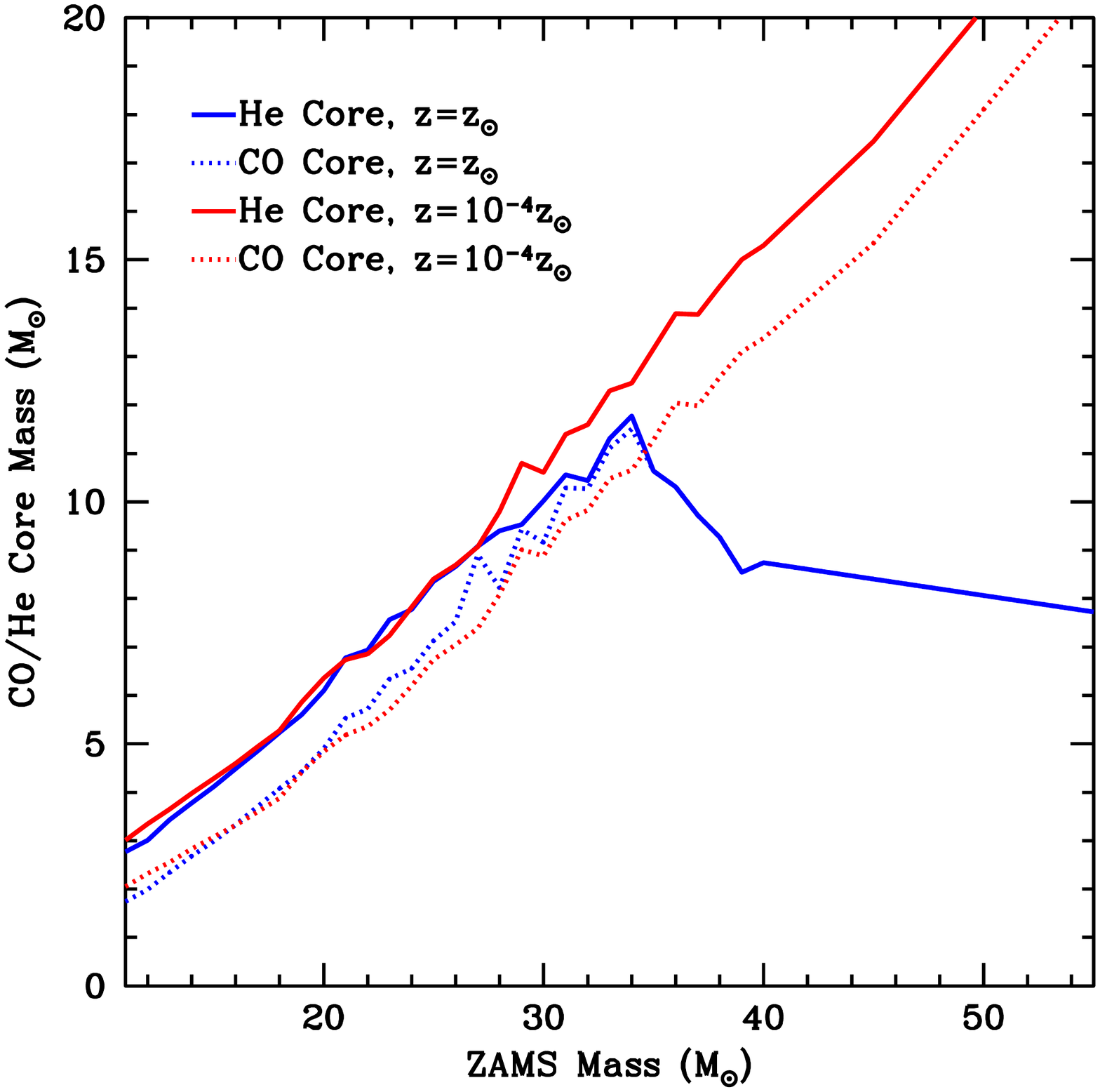}
\plotone{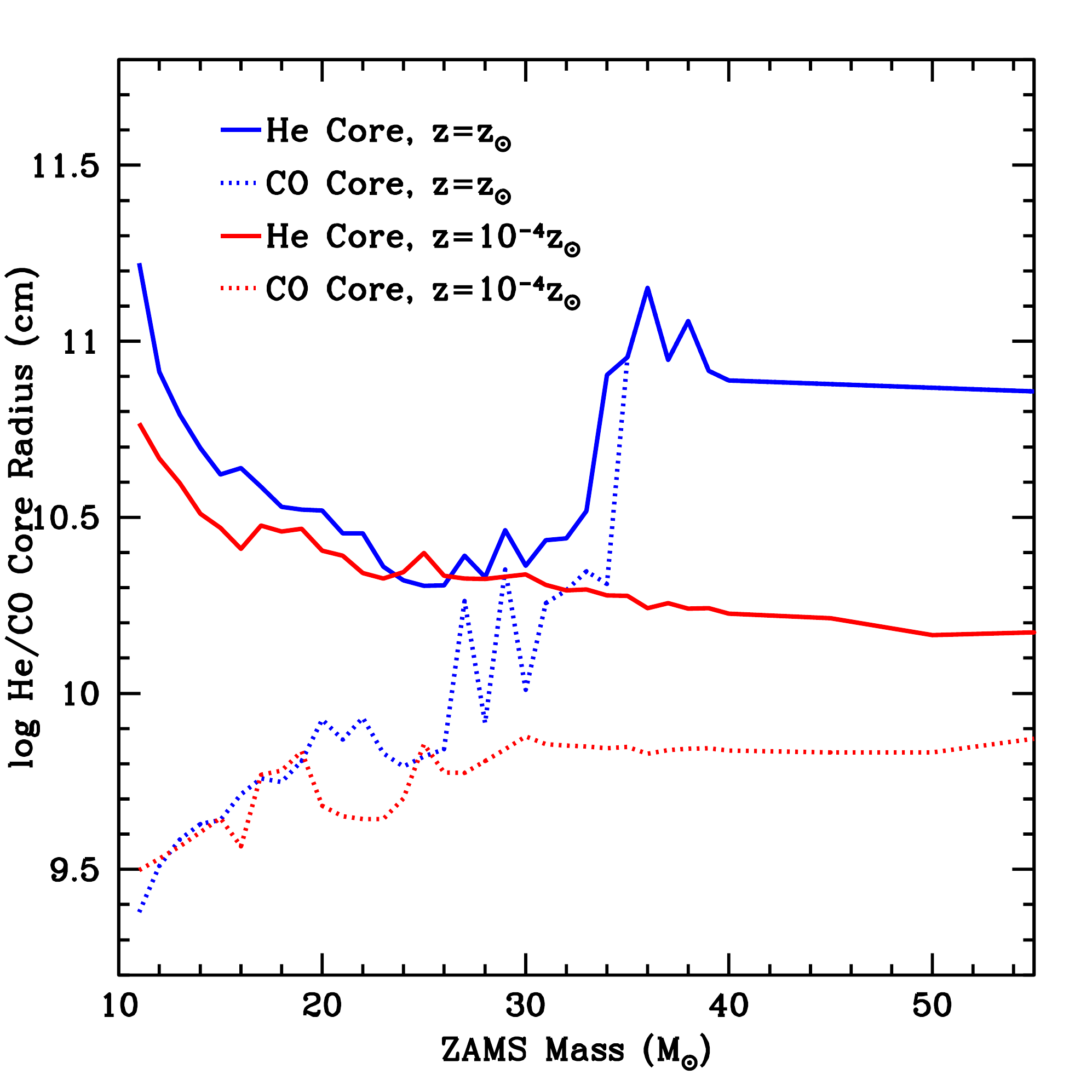}
\caption{Carbon/Oxygen and Helium core masses (top) and stellar radii (bottom) as a function of zero-age main sequence mass for stars at solar and $10^{-4}$ solar metallicities.  The primary difference between core masses is the mass loss for the most massive stellar progenitors.  There is a larger difference between the stellar radii.} % caption
\label{fig:core}
\end{figure}

While small compared to the effects at high metallicity, metallicity effects below 0.1\,z$_{\rm solar}$ can lead to detectable differences in the properties of a GRB.  This is because the structure of the stellar cores continue to change with metallicity, producing more compact cores that, as we shall discuss in section~\ref{sec:obs}, alter both the power in the GRB jet and the nature of the supernova produced along with this GRB.  For our study, we use the evolution of the progenitors from these GENEC models~\citep{2020AA...636A.104B}, using both the structure and angular momentum of these models as a function of metallicity.   Figure~\ref{fig:core} shows the radii and masses of the helium and Carbon/Oxygen cores as a function of zero-age main sequence (ZAMS) mass for a set of KEPLER models at solar and $10^{-4}$ solar metallicities~\citep{2002RvMP...74.1015W}. Note that although the core masses are roughly the same (or less massive at solar metallicity), the radii of the cores are typically higher at solar metallicity.

For our calculations, we have developed rough fits (as a function of mass and metallicity for these core masses and radii).  For the CO core mass, we use:
\begin{equation}
M_{\rm CO}=
\begin{cases}
    2.0+0.375 (M_{\rm ZAMS}-10),&
    \text{if } M_{\rm ZAMS} < M_{\rm crit} \\
    max(2.0+0.375 (M_{\rm ZAMS}- 10\,M_\odot)& \\
    - 0.1 (M_{\rm ZAMS}-M_{\rm crit}),2.0),&  \text{otherwise }
\end{cases}
\label{eq:mco}
\end{equation}
where $M_{\rm ZAMS}$ is the zero-age main-sequence mass of the star in M$_\odot$ and $M_{\rm crit}=180/(5+2log_{10}Z_{\rm metal})$.  For the corresponding radius of these CO cores, we use:
\begin{equation}
log_{10}(r)=
\begin{cases}
    9.5 + M_{\rm ZAMS}/100,& 
    \text{if } M_{\rm ZAMS} < M_{\rm crit} \\
    max(11,9.5+0.01M_{\rm crit}\\
    + 0.35(M_{\rm ZAMS}-M_{\rm crit})),&  \text{otherwise.}
\end{cases}
\label{eq:rzm}
\end{equation}

\subsubsection{Metallicity Distributions with Redshift}
\label{sec:zvz}

For our metallicity-redshift relation, we use the same prescriptions used by \cite{2007ApJ...670..584Y,2013ApJ...779...72D}.  For the mean metallicity with redshift, we use the value described by \cite{1999ApJ...522..604P}:
\begin{equation}
Z_{\rm metal}=
\begin{cases}
    10^{-a_2 z},& \text{if } z<3.2 \\
    10^{-b_1 - b_2z},& \text{if } 3.2<z<5 \\ 10^{-c_1 - c_2z},&  \text{if } z>5
\end{cases}
\label{eq:zm}
\end{equation}
where we use the fitting parameters set by~\cite{1999ApJ...522..604P}:  $a_2=0.5, b_1=0.8, b_2=0.25, c_1=0.2, c_2=0.4$.
%\textcolor{red}{We should probably vary these.  %Another set has:$a_2=0.12, b_1=-0.704, b_2=0.34, %c_1=0.0, c_2=0.1992$
%\citep{2006ApJ...651..142H,2007ApJ...670..584Y}.}

We allow for a distribution of metallicities as a function of redshift by assuming different galaxy masses have different metallicities~\citep{2004ApJ...613..898T}:
\begin{eqnarray}
    12+log(O/H) &=& -1.492 + 1.847 log M_{\rm galaxy} \\ &-& 0.08026 (log M_{\rm galaxy})^2
\end{eqnarray}
where the metallicity is set be the relative abundance of Oxygen to Hydrogen.  For the galaxy mass distribution with redshift, we again use the formalism by ~\cite{2007ApJ...670..584Y,2013ApJ...779...72D} using the fitting formulae from~\cite{2006AA...459..745F}:
\begin{equation}
    f(M_{\rm galaxy}) \propto log \left(10^{log(M)-M_*(z)}\right)^{\alpha_*}(z) exp({10^{log(M)-M_*(z)}})
\end{equation}
where $log(M)$ is the logarithm of the galaxy mass in units of solar masses, $M_*(z)=11.16+0.17z-0.07z^2$, and $\alpha_*(z) = -0.18-0.082z$.  Above a redshift of 4, we set the mass function to the value at a redshift of 4.

\section{Redshift Evolution of Observed Transients}
\label{sec:obs}

In section~\ref{sec:progenitor}, we discussed the evolution in massive star structures and populations as we move to higher redshifts and lower metallicities.  These structural properties will imprint themselves on the properties of the explosion produced from gamma-ray bursts and their associated supernovae.  In this section we discuss the evolution on these observed transient properties.

To do so, we must assume properties of our GRB model and its progenitors.  Although a wide set of GRB models exist, only the Black-Hole Accretion Disk (BHAD) models~\citep{1999ApJ...526..152F} have successfully made predictions for the observed GRBs~\citep{2019EPJA...55..132F}. These predictions include the differentiation of the progenitors of short and long bursts:  e.g. short bursts are binary mergers arguing for an offset of the bursts with host galaxy and its star forming regions and long bursts are produced from very massive stars and correlated with robust star forming regions~\citep{1999ApJ...526..152F,1999MNRAS.305..763B} and, because massive star progenitors have strong winds, a dearth of GRBs near solar metallicity~\citep{2007ApJ...670..584Y}.  Magnetar and neutron star accretion disk models have not explained these GRB features.  With this in mind, we will focus on black hole accretion disk models.  For this study, we will also study only the properties of long-duration gamma-ray bursts that tie most-directly to the formation of massive stars.

Having assumed BHAD models, we still have to make assumptions on the nature of the progenitor of these models.  Two aspects of the GRB paradigm are currently not completely understood by the existing progenitors:  1) how to produce enough angular momentum to make an accretion disk around the black hole and 2) the observational feature that, thusfar, it appears that the supernovae associated with GRBs are type Ic~\citep{2007PASP..119.1211F}\footnote{\cite{2014ApJ...793L..36F} argue that the star must undergo both a hydrogen and helium common envelope, explaining the fact that GRB-associated supernovae are type Ic.}.  Both of these features point to either binary interactions~\citep{1999ApJ...526..152F,2007ApJ...670..584Y} or extensive stellar mixing~\citep{2013ApJ...773L...7F}.

Regardless of the exact progenitor, we will assume the trends with metallicity in the structures of single star models persist for the progenitors of GRBs.  We assume that the properties of GRBs will scale with the properties of metallicity dependence of single stars.  With these structural changes and the IMF dependencies, we can study the evolution of the GRB and GRB-associate supernovae properties (peak luminosity and durations) with redshift.   We also include the evolution of the IMF and interacting binary fraction in studying rates.

\subsection{GRB Durations and Luminosities}
\label{sec:lgrb}

By assuming that GRB observables trace the single-star evolution, we are effectively studying collapsar progenitors for different collapsar types~\citep{2003ApJ...591..288H}:  
\begin{itemize}
    \item Type I collapsars are normal GRBs produced by massive stars that form direct-collapse black holes.  These systems do form a proto-neutron star, but the the standard convective engine is unable to produce a supernova explosion.  The GRB-associated supernova is produced by the GRB itself.  This is our standard GRB model.
    \item Type II collapsars are GRBs produced by stars who first produced weak SN explosion.  The subsequent fallback produces a black hole.  We will assume these systems make dim GRBs because we expect their accretion rates to be lower than the accretion rates of Type I collapsars.  These bursts will only be detected at low redshift and we do not include these systems in the study of high redshift GRBs.
    \item Type III collapsars are GRBs produced in the most massive stars (above the pair-instability gap).  As we shall discuss below, these collapsars may be much dimmer than normal GRBs because these systems collapse from large proto-black holes~\citep{2001ApJ...550..372F} and engine weakens significantly with increasing black hole mass.
\end{itemize}

\cite{1999ApJ...518..356P} estimated the strength of a Blandford-Znajek jet based on the properties of  the black hole and accretion disk.  In this paradigm, the magnetic field develops in the disk and its strength is relative to the disk energy.  This is directly proportional to the accretion rate, but also depends on the spin and mass of the black hole.  We use the following fit to the data from \cite{2003ApJ...591..288H}:
\begin{align}
L_{\rm BZ} =  & f_{\rm Lpeak} 10^{50} a_{\rm spin}^2 10^{0.1/(1-a_{\rm spin})-0.1} \nonumber \\
& \left( \frac{M_{\rm BH}}{3 M_\odot} \right)^{-3} \frac{dM/dt}{0.1M_\odot/s} {\rm \, erg \, s^{-1}}
\label{eq:lgrb}
\end{align}
where $M_{\rm BH}, a_{\rm spin}$ are the black hole mass and spin respectively, $dM/dt$ is the accretion rate, and $f_{\rm Lpeak}$ is a parameter that covers the efficiency at which disk energy is converted to the jet power and, ultimately, the GRB luminosity.  

\begin{figure}
\plotone{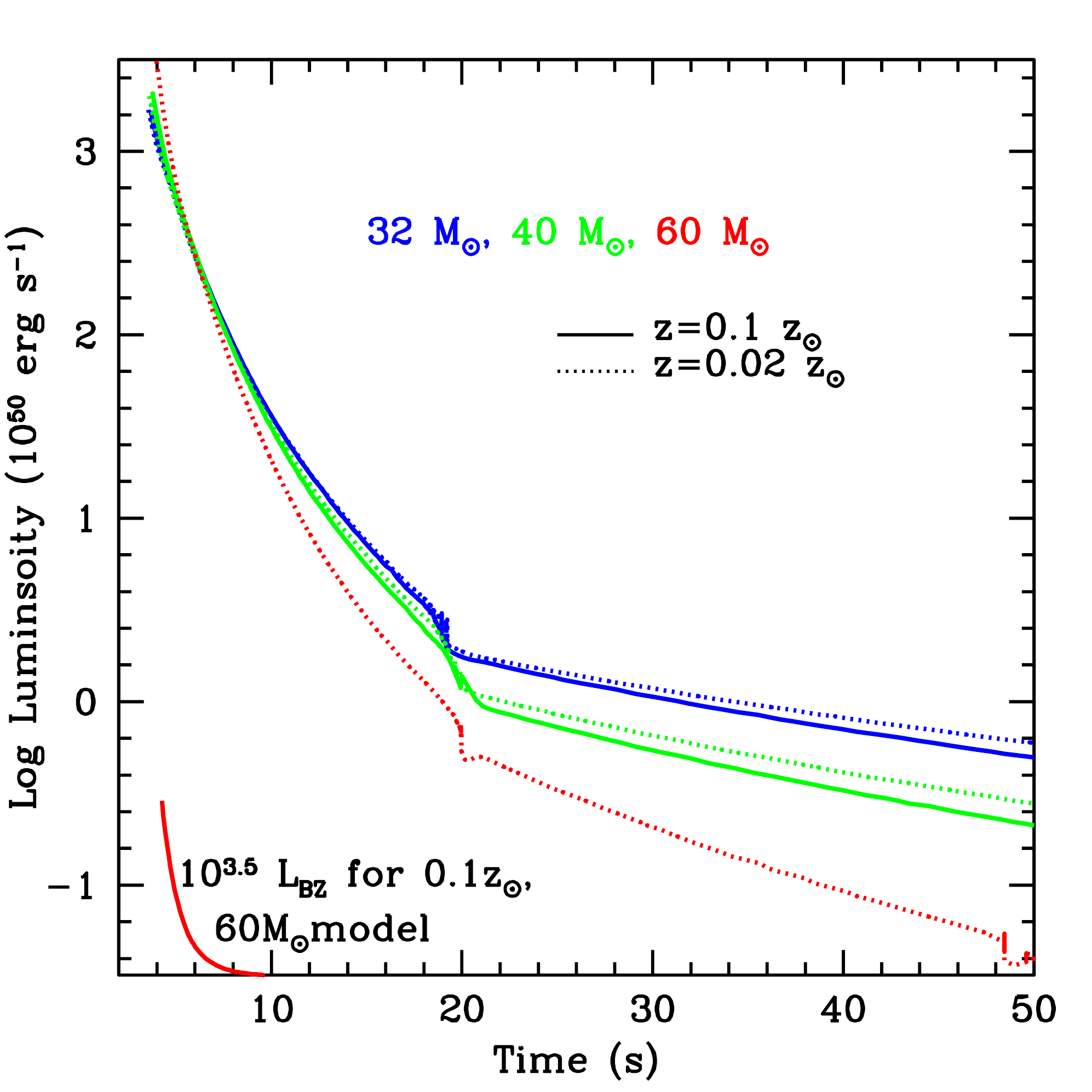}
\caption{Log GRB jet luminosity as a function of time for a range of progenitors masses at 2 metallicities:  0.01 and 0.02 solar metallicity.   Note that more massive type II collapsar progenitors produce more luminous GRBs.  As the IMF flattens, the mean burst luminosity (with more contributions from more massive stars) will increase.  There is a slight trend where the peak luminosity increases with decreasing metallicity, but this effect is small and is negligible below 0.02\,z$_\odot$.
} % caption
\label{fig:grblum}
\end{figure}

With this formula, we can derive the power in the jet and, if we assume there is a one-to-one correspondence between jet power and GRB luminosity (assuming some efficiency less than 1), the GRB luminosity as a function of time (Figure~\ref{fig:grblum}).  There is a generic trend in the luminosity where the power in the jet decreases with time as the black hole accretes mass and the event horizon moves outward.   Another feature of the models is that, at high metallicities, the more massive stars can not drive strong jets.  This is because mass loss removes the angular momentum of the star and the low black hole spin limits the power from the Blandford-Znajek mechanism.   

\begin{figure}
\plotone{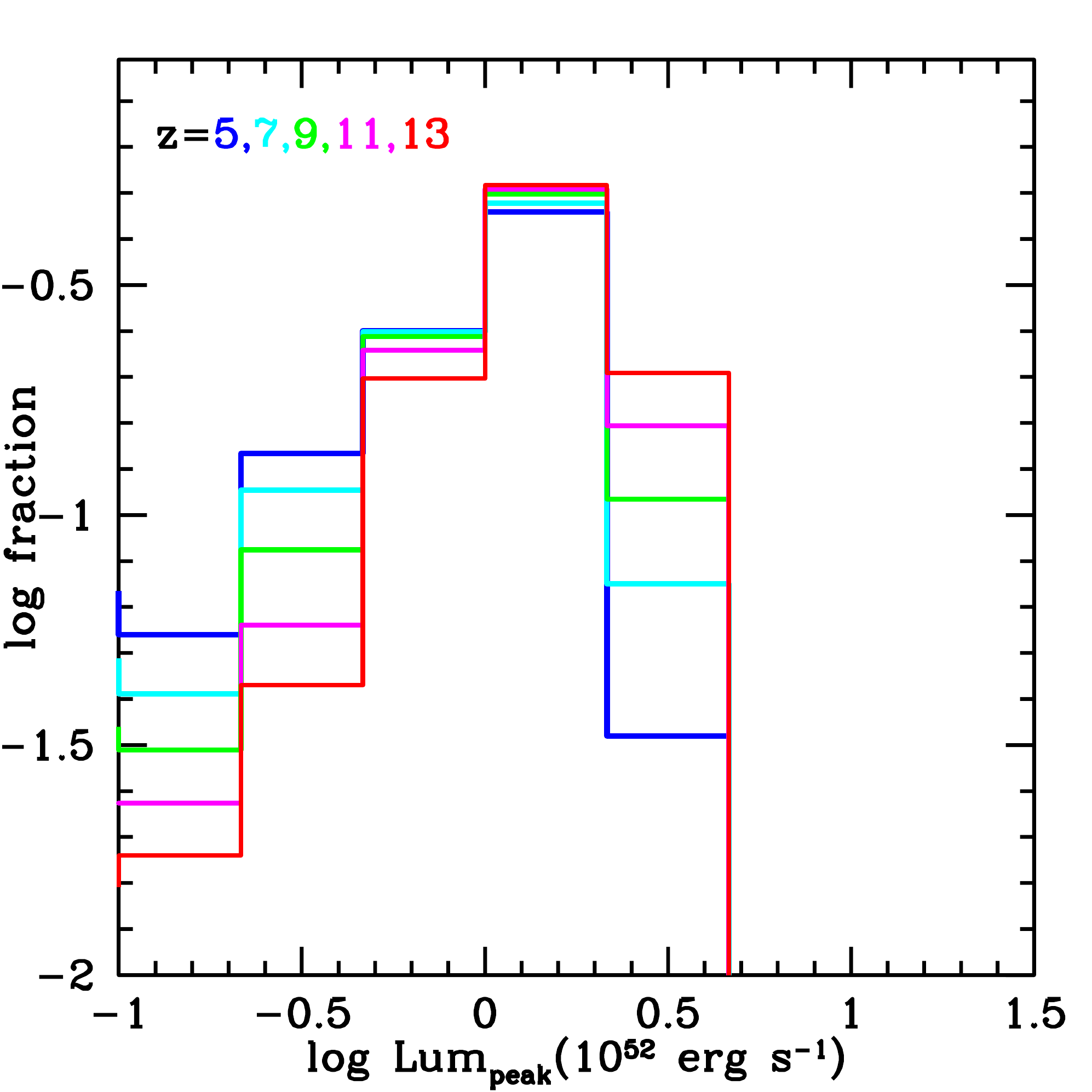}
\plotone{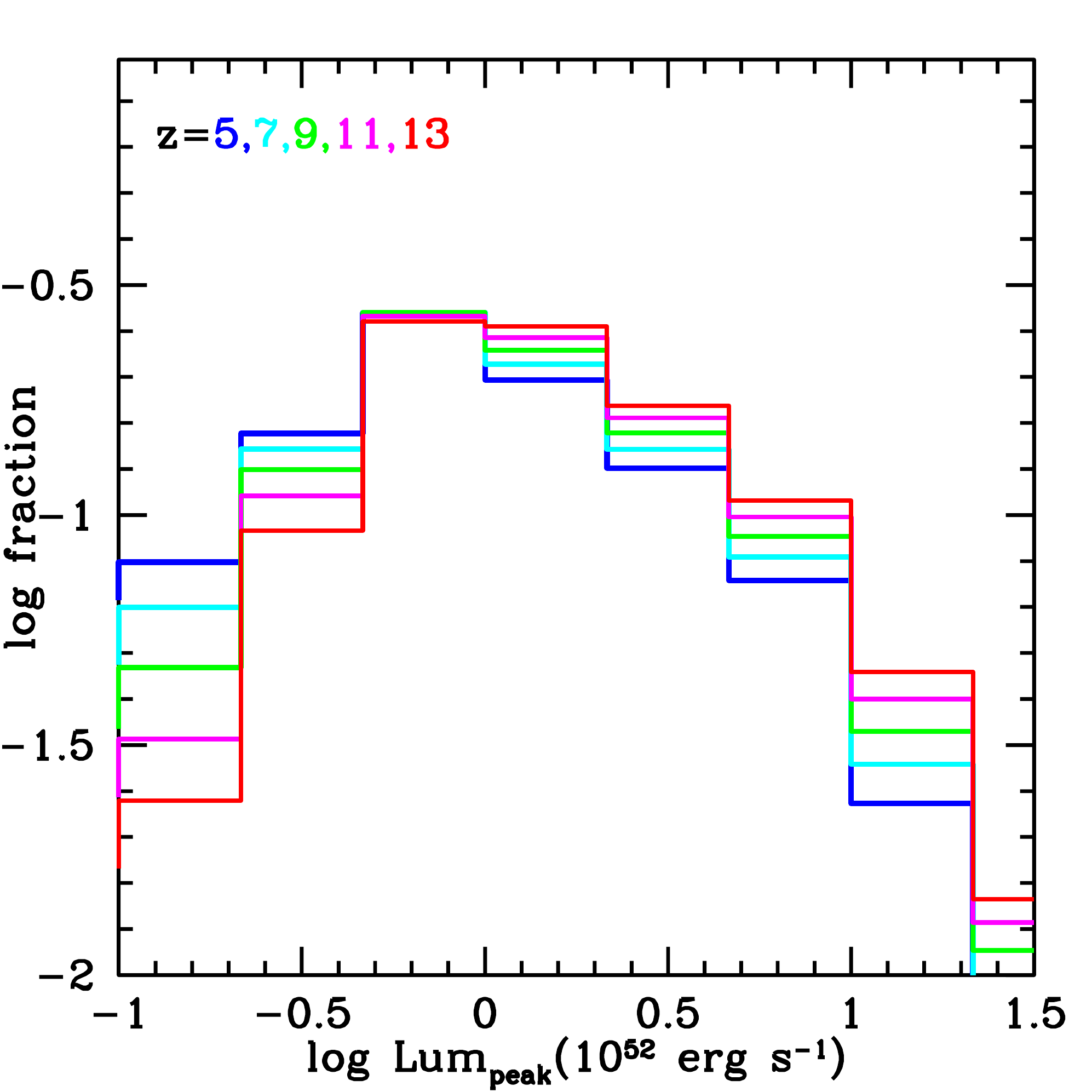}
\caption{Top:  Distribution of peak GRB luminosities at 5 different redshifts for our $f_{\rm IMF}=0.4, z_0=5, f_{\rm bin}=0$ IMF evolution model.  As the IMF flattens out, the GRBs become more luminous, producing slightly more high-luminosity GRBs.  Bottom:  Distribution of GRB luminosities at 5 different redshifts for the same model
where a spread in the GRB luminosities is included.} 
\label{fig:grblumdist}
\end{figure}

As we move to high redshift, the metallicity decreases (more massive stars don't lose their angular momentum) and the IMF flattens out.  If we include both effects, using our metallicity vs. redshift relationship, we can study the evolution of the distribution of GRB luminosities as a function of redshift.  Figure~\ref{fig:grblumdist} shows these peak luminosity distributions as a function of redshift.  In one our $f_{\rm IMF}=0.4, z_0=5, f_{\rm bin}=0$ model, the IMF flattens producing a larger number of bright GRBs. 

Variations in the spin rate and initial seed magnetic fields between different stars should broaden this distribution.  Without a better theoretical understanding of stellar evolution, this is difficult to predict.  We study these effects (see Section~\ref{sec:dim_GRBs}) by assuming that our estimates produce average luminosities and using observed distributions as a guide~\citep{2012ApJ...749...68S,2015MNRAS.447.1911P}.  These observational studies suggest a distribution $f(L) \propto (L/L_{\rm break})^{-a}$ where $a=1.5,2.3$ below and above the break luminosity ($L_{\rm  break}$).  In this distribution model, our nominal luminosity corresponds to the "break" luminosity in these studies and the variation in the luminosity with redshift becomes even less pronounced.  Although these observations also predict a large number of dim bursts (which may be caused by lower-mass progenitors with fallback black holes), our focus in our study is on bright GRBs and we limit our distribution to these bright bursts.

If the level of beaming of GRBs depends on the mass of the progenitor, a changing IMF may also change the level of beaming.  At this time, there is no clear theoretical argument to suggest a variation in beaming.  Although it is possible that different disk angular momenta or stellar structures can cause a variation in the beaming, we have no models showing how these will affect the jet beaming.  Most of the stellar-evolution differences from metallicity evolution occur at low redshift (below redshift 4), i.e., we do not expect an evolution in beaming at high redshift for a given-mass progenitor.  But, if different-mass stars produce different beaming factors (potentially observable at lower redshift), an evolving IMF could alter the beaming.  Some observations suggest a mild beaming dependence on the GRB luminosity:  $\Theta_{\rm beaming} \propto L_{\rm GRB}^{0.2}$\citep{2007A&A...466..127G} which corresponds to a small difference in opening angle with redshift.  Although some analyses argue for a stronger effect~\cite{2020MNRAS.498.5041L}, without a strong theoretical argument for beaming, most of our models do not include beaming evolution.  If GRBs become more beamed with redshift, we would expect a lower GRB rate with redshift than the results from our models without beaming evolution.

%It is more difficult to estimate the distribution of dim GRB luminosities produced from fallback black holes and we will not estimate this redshift evolution in this paper but assume that the dim GRB luminosity distribution remains constant.  Given that these dim bursts will be difficult to observe at high redshift, assuming no evolution in the peak luminosity of dim GRBs does not affect our high-redshift GRB observations.

\begin{figure}
\plotone{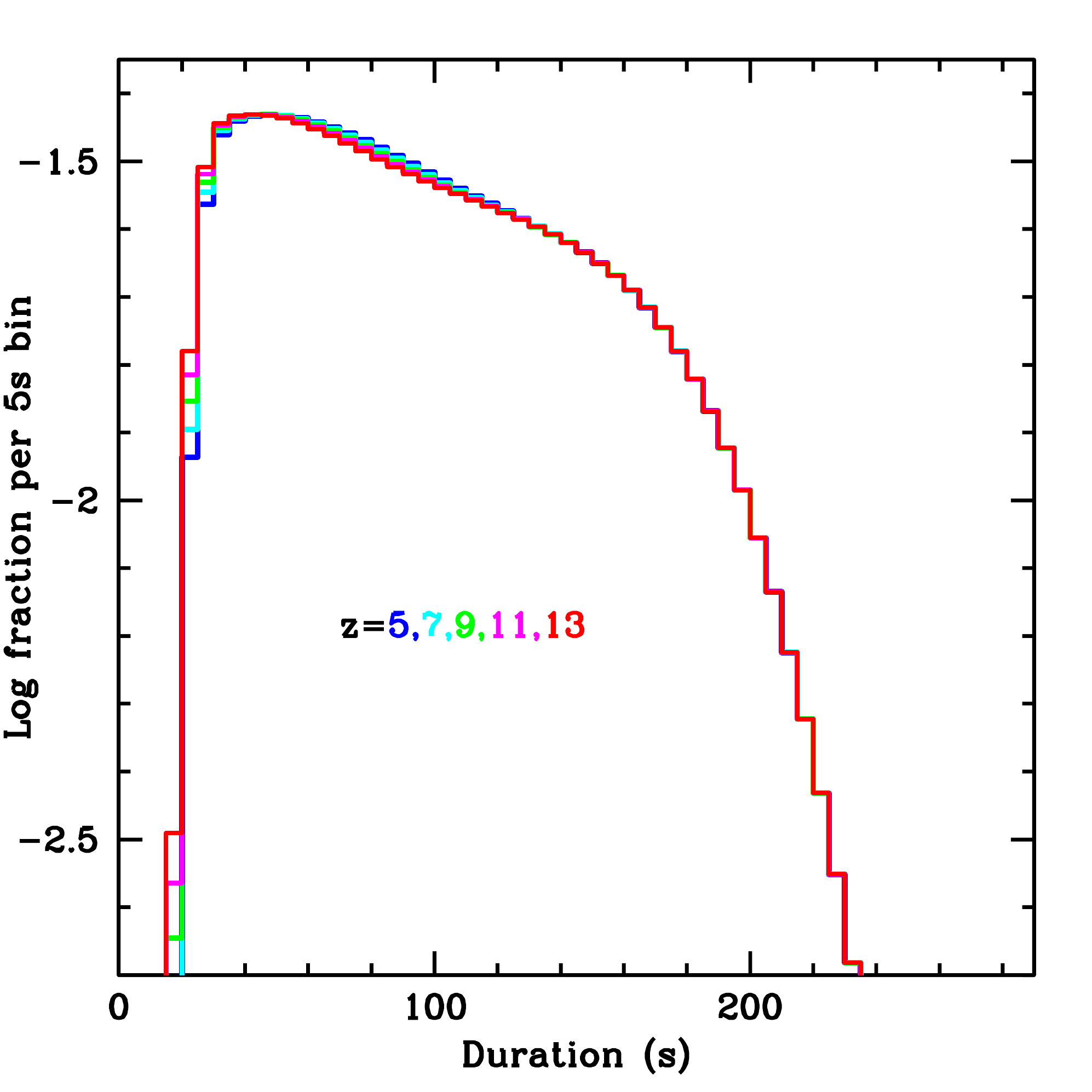}
\caption{Distribution of the durations of the GRB power source for a range of redshifts.  Because of the strong dependence of the GRB luminosity on the black hole mass, the duration does not evolve considerably with mass (see Fig.~\ref{fig:grblum}).
} % caption
\label{fig:tdur}
\end{figure}

We can estimate the duration as well based on the time it takes for the GRB jet luminosity to drop one order of magnitude.   The primary constraint on the duration of the GRB is the accretion rate.  If the black hole accretes mass quickly, its event horizon will move outward, reducing the energy in the jet under this disk-driven jet paradigm.  What this means is that, even if the mass accretion rate remains steady for long periods of time, the GRB duration will drop off.  As with the GRB luminosity, the difficulty in calculating the duration is tying the power source of this jet to a GRB duration.  Figure~\ref{fig:tdur} shows the distribution of the duration of the power source (timescale for the luminosity to drop by an order of magnitude) as a function of redshift.  Because of the strong dependence of the luminosity on the black hole mass, there is very little evolution in this duration.  

\subsection{Ibc Supernova evolution}

A supernova-like transient is also produced when the GRB jet disrupts the star.  It drives a strong, asymmetric explosion called a hypernova\citep{1998Natur.395..672I}.  Just as the structure of the star alters the conditions around a black hole accretion disk, the compactness of the core can alter the transient produced when the jet disrupts the star.  To first order, type Ic hypernovae rely on the same physics (transport of energy from a $^{56}$Ni power source - but note that shock heating may also be important) as type Ia supernovae and can be understood from the physical understanding from studies of those supernovae.  The Arnett law describes the peak luminosity as~\citep{1980ApJ...237..541A}:
\begin{equation}
    L = 2.055 \times 10^{10} L_{\odot} M_{\rm Ni} \Lambda(x,y)
\end{equation}
where $M_{\rm Ni}$ is the nickel mass and $\Lambda(x,y)$ is a dimensionless function that depends upon a shaping parameter $y=(\kappa M_{\rm ejecta}/v_{\rm explosion})^{1/2}$ where $\kappa, M_{\rm ejecta}$ and $v_{\rm ejecta}$ are the opacity, mass and velocity of the ejecta, .  Under this formalism, the peak luminosity depends on the following properties of the explosion:  ejecta mass, the nickel mass, the composition (for the opacity), and the explosion energy ($E_{\rm exp}$).

\begin{figure}
\plotone{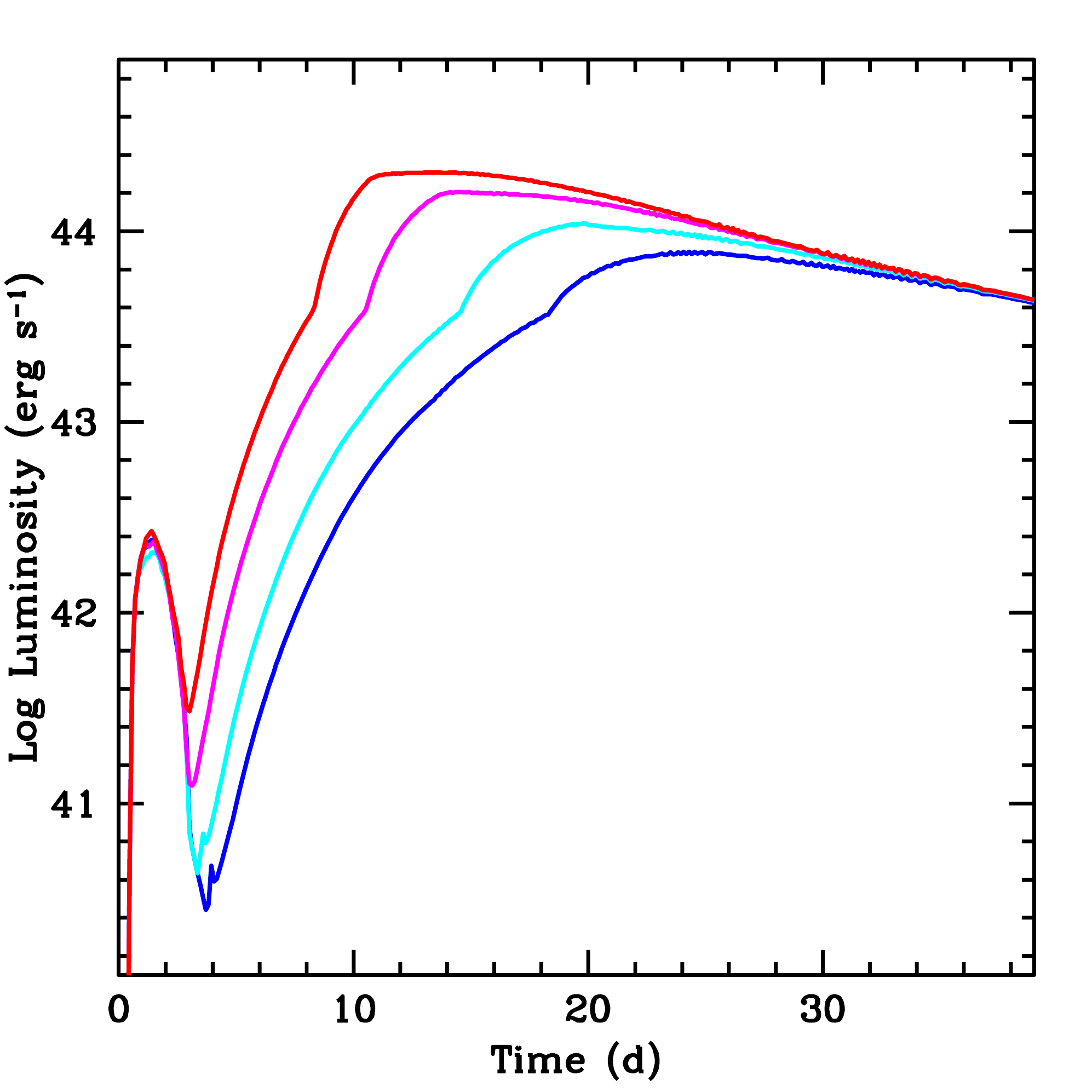}
\plotone{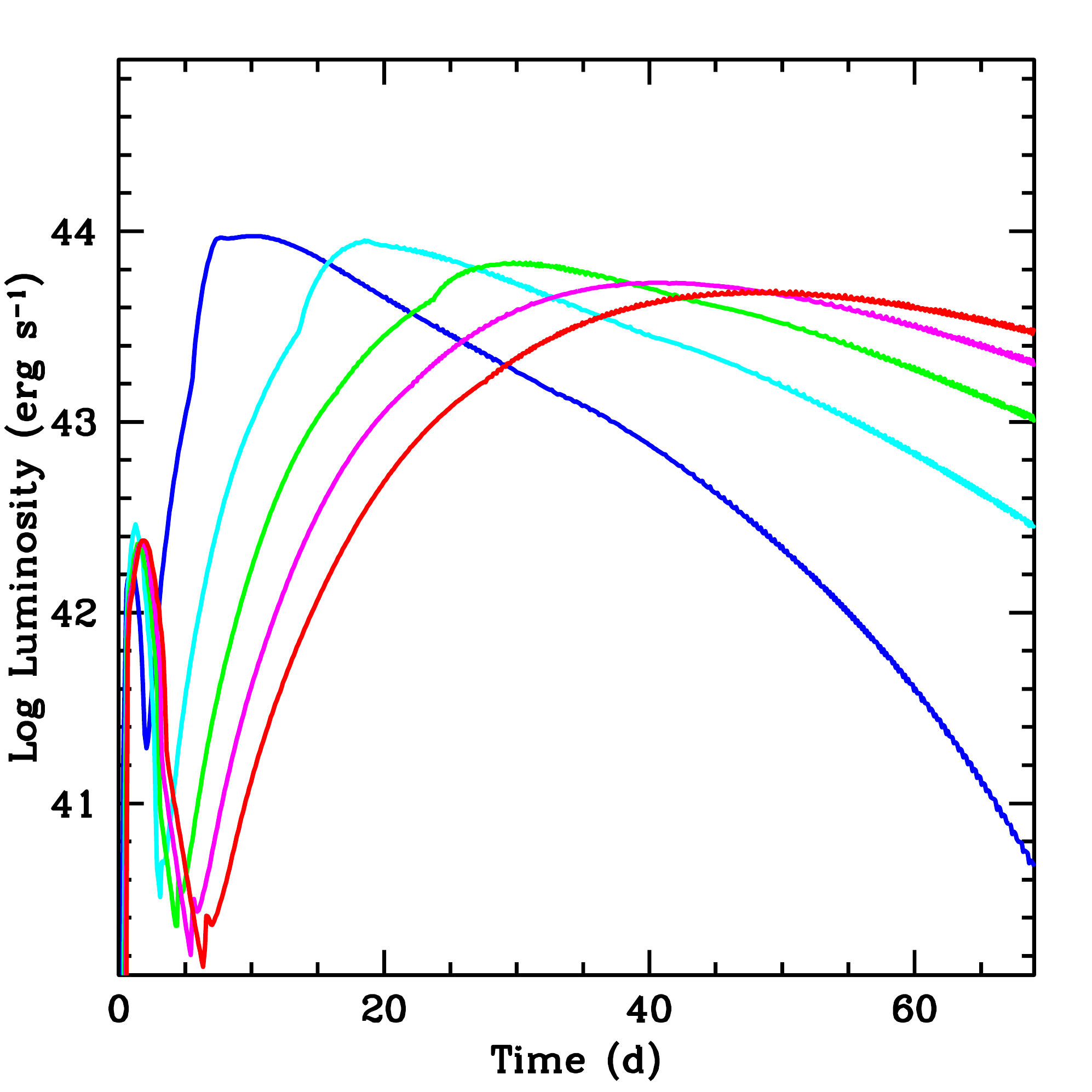}
\plotone{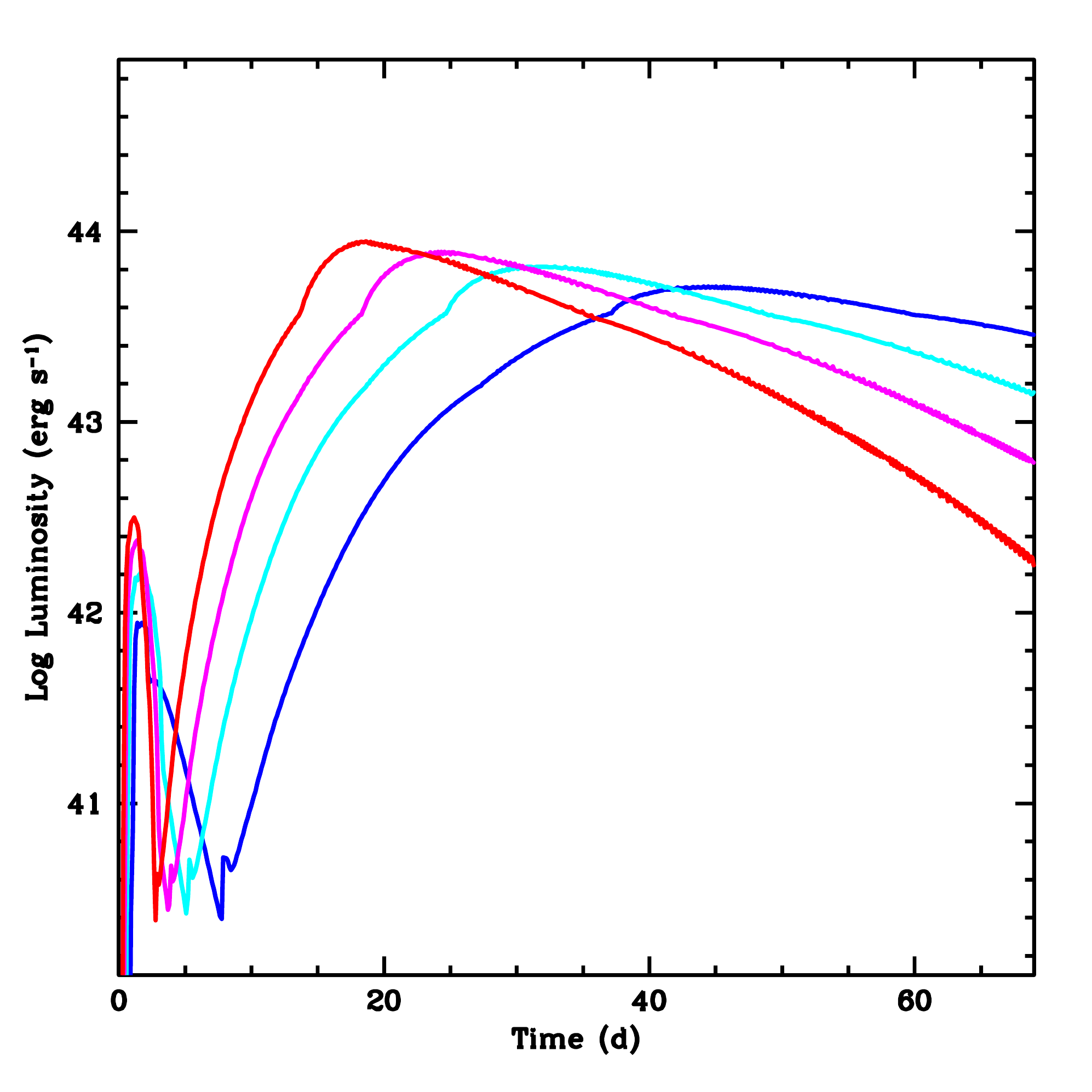}
\caption{Luminosity versus time for type Ic light-curves varying the nickel mass (top), ejecta mass (middle), explosion energy (bottom).} % caption
\label{fig:ibcdepend}
\end{figure}

To understand these dependencies, we have run a series of calculations of supernovae varying the ejecta mass, nickel mass, explosion energy and radius.  For these calculations, we use the gray diffusion supernova code described in \cite{2017ApJ...850..133D}, designed for large grids of supernova calculations.  It includes a flexible framework for initial conditions (to probe the dependencies on the progenitor and explosion properties) and a recipe to include shock heating.  Figure~\ref{fig:ibcdepend} presents the results of some of these explosion, showing trends in nickel and ejecta masses as well as explosion energy.  From these calculations, we have derived fits to the peak luminosity and duration (defined as the time spent within one e-folding of the peak luminosity):
\begin{equation}
    L_{\rm peak} \propto M_{\rm ejecta}^{-0.43} M_{\rm Ni} E_{\rm exp}^{0.024}
    \label{eq:libc}
\end{equation}
where $M_{\rm ejecta}$ is the ejecta mass, $ M_{\rm Ni}$ is the nickel mass and $E_{\rm exp}$ is the explosion energy.  The corresponding 
\begin{equation}
    t_{\rm dur} \propto M_{\rm ejecta}^{0.68} M_{\rm Ni}^{-1.25} E_{\rm exp}^{0.032}.
    \label{eq:tibc}
\end{equation}

\begin{figure}
\plotone{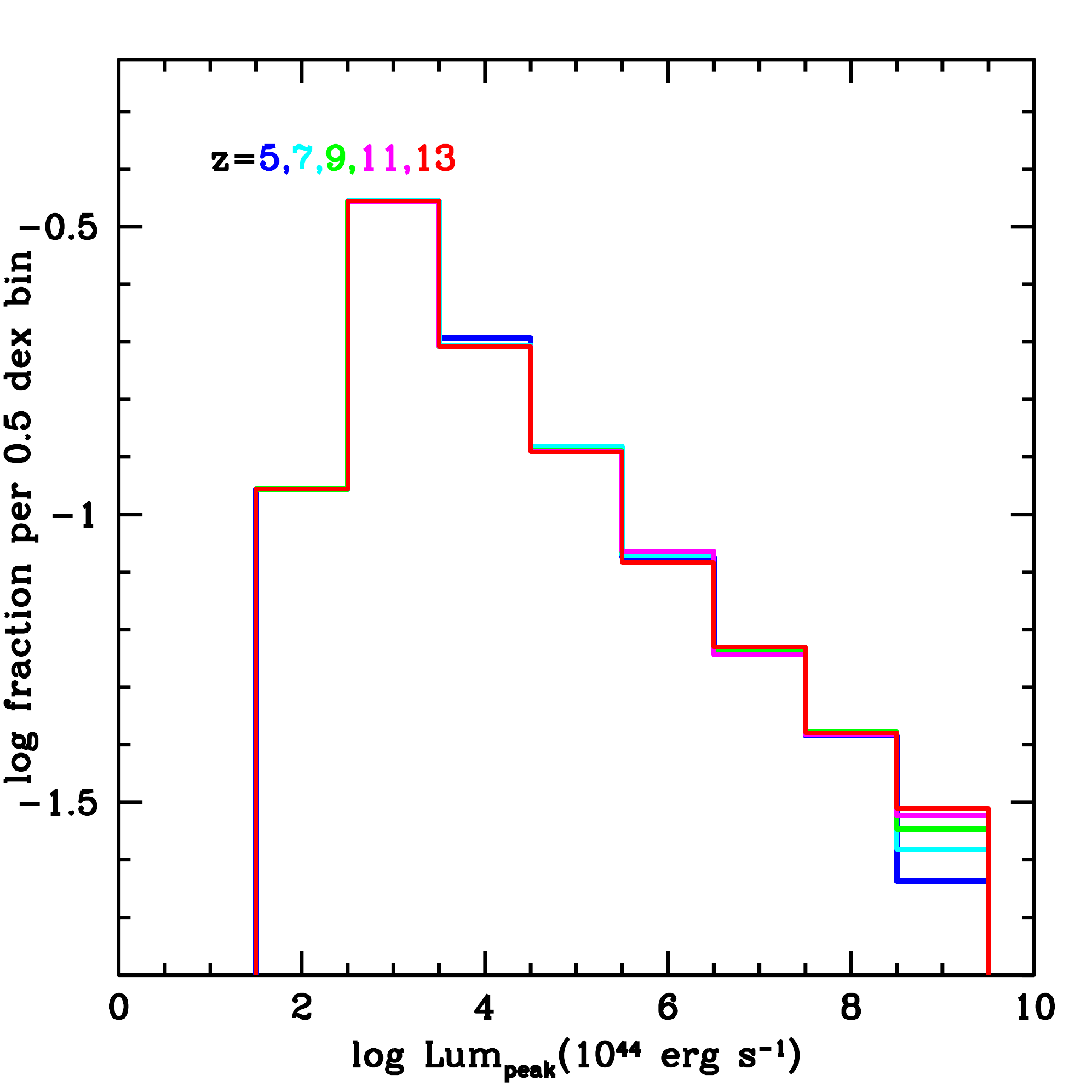}
\plotone{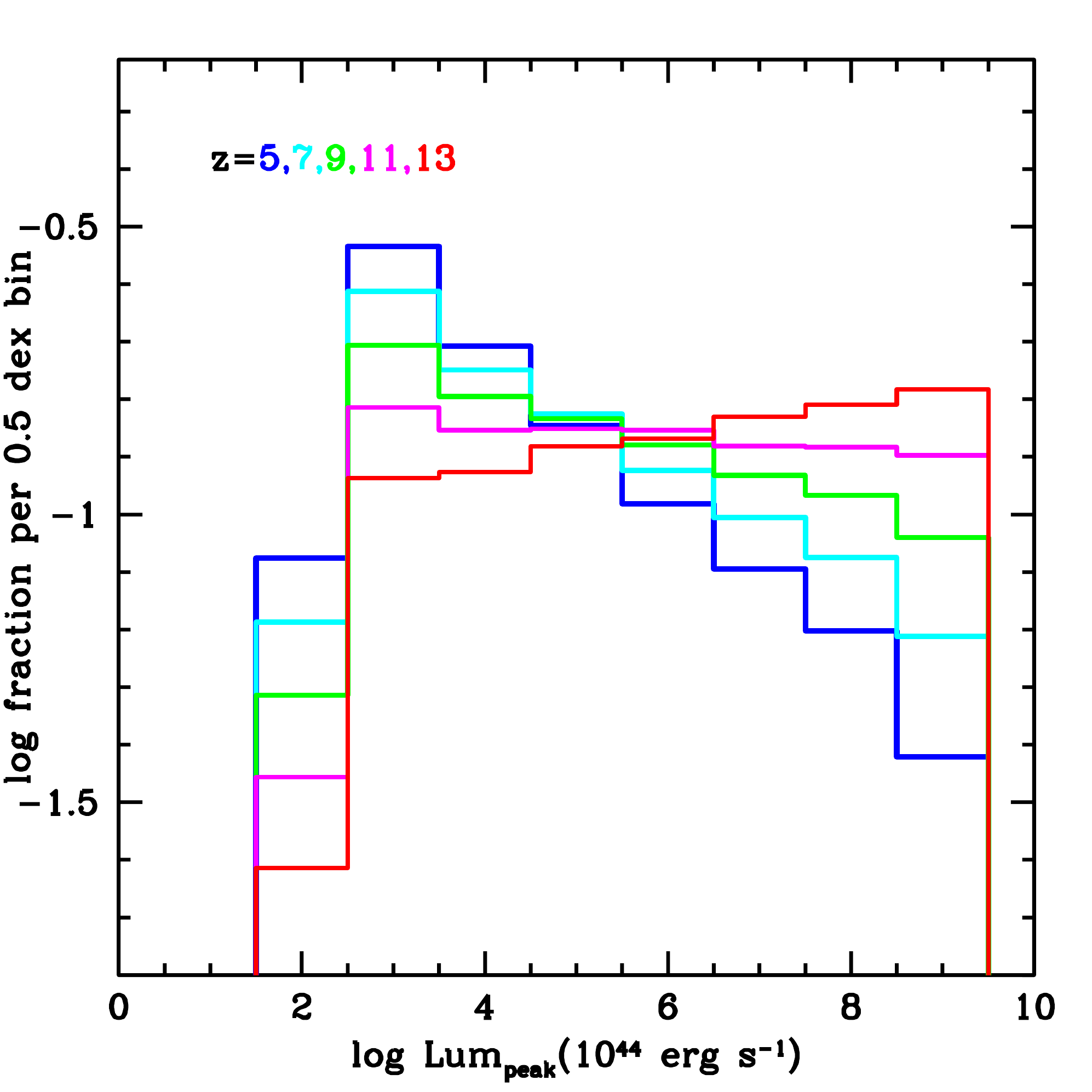}
\caption{Distribution of Ibc luminosities as a function of redshift for a model where the IMF does not evolve with redshift (top) and a model where the IMF changes with redshift (bottom):  $f_{\rm IMF}=0.2, z_0=2.5$.} % caption
\label{fig:libcdist}
\end{figure}

\begin{figure}
\plotone{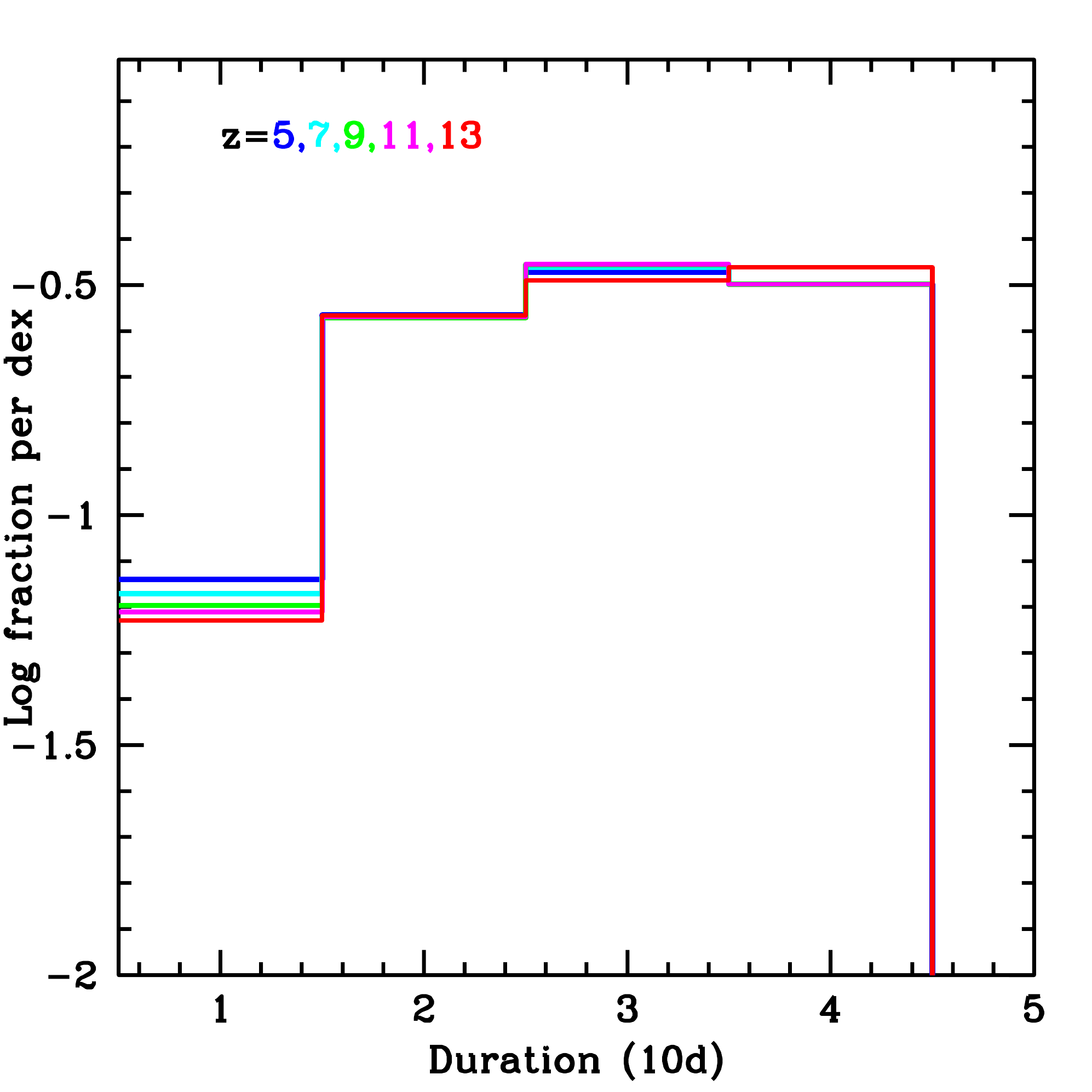}
\plotone{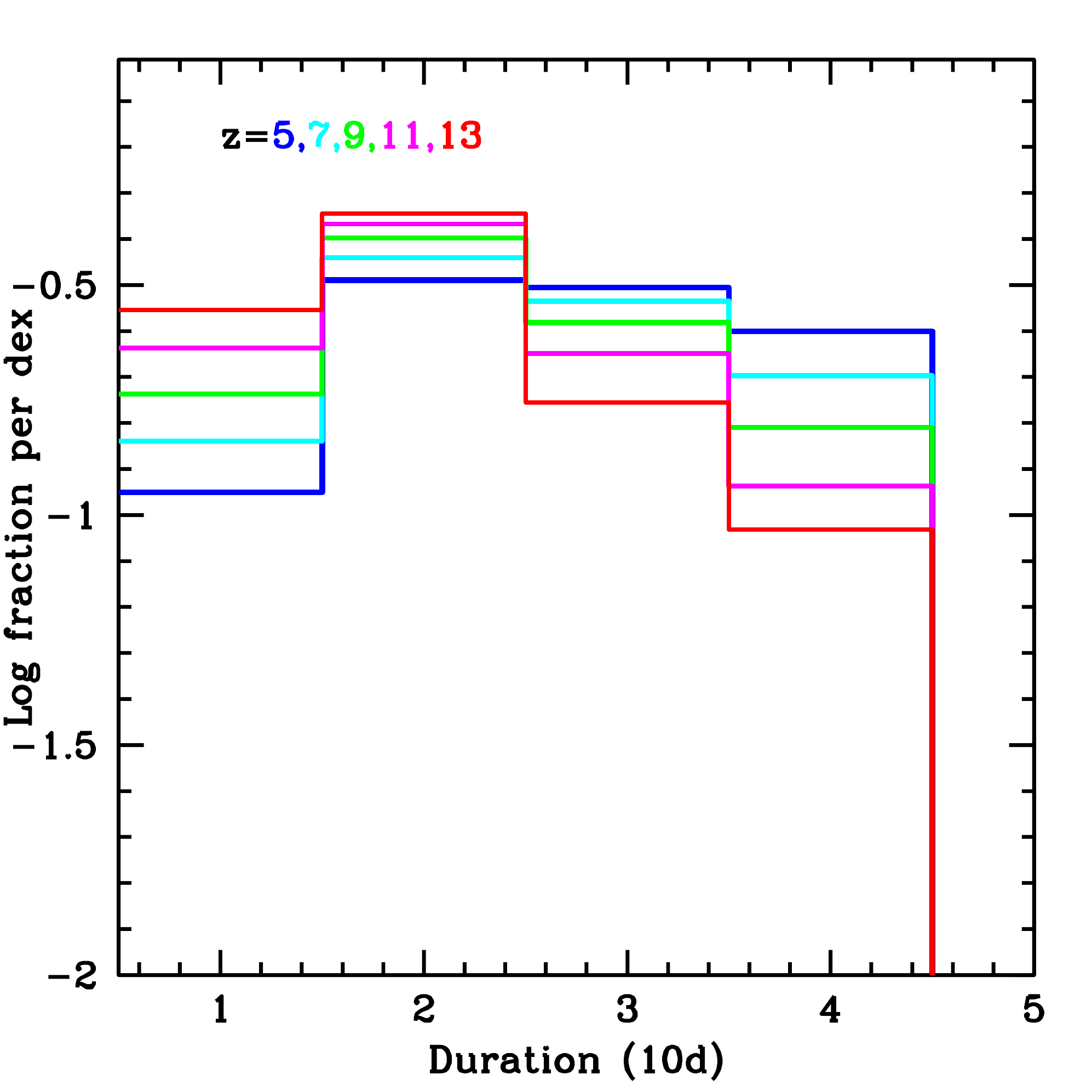}
\caption{Distribution of Ibc durations as a function of redshift for a model where the IMF does not evolve with redshift (top) and a model where the IMF changes with redshift (bottom):  $f_{\rm IMF}=0.2, z_0=2.5$.} % caption
\label{fig:tibcdist}
\end{figure}

The nickel production in a collapsar is either produced in the disk wind~\citep{2004ApJ...603..611S} or in the shock as ejecta plows through the star~\citep{2001ApJ...555..880N}.  In both cases, it depends on the density, and hence mass, of the core.  For a  zeroth-order approximation of this dependence, we have assumed that the nickel mass, $M_{\rm Ni}$, is proportional to CO core mass, $M_{\rm CO}$.  By combining this assumption to our fitted formulae for the supernova peak luminosity and duration (equations~\ref{eq:libc},\ref{eq:tibc}), our supernova properties depend only on the CO core mass and the explosion energy.  Using our stellar and explosion properties discussed earlier in the paper (equations~\ref{eq:mco},\ref{eq:lgrb}), the metallicity dependence on redshift (equation~\ref{eq:zm}), and the IMF dependence on redshift (equation~\ref{eq:imf}), we can model the distribution of both peak luminosities and supernova durations.  Figure~\ref{fig:libcdist} shows the distribution of peak supernova luminosities as a function of redshift for two different models from the evolution of the IMF.  The corresponding distributions for the supernova durations (defined by the time when the luminosity is within a e-folding of the peak flux) are shown in figure~\ref{fig:tibcdist}.

Although this discussion focused on GRB-associated hypernovae, we note that hypernovae may also be produced in systems where the jet fails to produce a strong burst of gamma-rays (i.e. the jet decelerates before exiting the star).  The rate of hypernovae is difficult to determine.  \cite{Guetta_2007} found that roughly 7\% of all SNIb/c exhibited broad lines characteristic of hypernovae and that the observations of hypernovae suggested a relative rate of normal GRBs to hypernovae is somewhere between 4-30\%.  This difference could be explained simply by the beaming fraction of GRBs and \cite{Guetta_2007} argued thatt the hypernova rate was consistent with the normal GRB rate.  For our models, we assume there are twice as many hypernovae as GRBs (presumably formed from dim GRBs or failed-jet hypernovae).  

\subsection{Pair-Instability Supernovae}

Other types of massive star explosions also evolve with a changing IMF and by combining GRB results with observations of other powerful explosions, we can further constrain the IMF~\cite{2021arXiv211011956L}.  Most notable are the pair-instabiliy supernovae produced in a runaway thermonuclear burning phase instigated by electron/positron pair production in a hot core~\citep{1967PhRvL..18..379B}.  Although it appears the exact stellar masses that undergo pair-instability is sensitive to the $^{12}C(\alpha,\gamma)$ rate~\citep{2020ApJ...902L..36F,2020arXiv201002242C}, we use the limits and metallicity dependencies set by \cite{2001ApJ...550..372F,2003ApJ...591..288H}.  At high metallicities, winds reduce the stellar mass, preventing the high temperatures needed to make pairs.  We include these mass loss effects reducing the rate of pair-instability supernovae, but we assume that, if an explosion is produced, its yields and luminosity do not change.  

\section{Calibrating and Comparing Models to Current Data}
\label{sec:calibration}

There are a number of uncertainties in our simple model: e.g, the efficiency factor in energy conversion to produce the GRB luminosity ($f_{\rm Lpeak}$) in equation~\ref{eq:lgrb}, the role of mass loss from winds ($f_{\rm wind}$) in equation~\ref{eq:wind}, and the fraction of dim bursts (and their related hypernovae).  The rate of GRBs is determined both by the number of systems with sufficient angular momentum to form a disk ($f_{\rm disk}$) and the beaming fraction of the GRB jet ($f_{\rm beam}$).  For our rate study, these two uncertainties are degenerate and we can use a single rate parameter in our calibration: $f_{\rm GRB} = (f_{\rm disk}/0.01) \times (f_{\rm beam}/0.01)$.  In other words, $f_{\rm GRB}$ is the fraction of stars with sufficient angular momentum to produce GRBs and with a beaming angle that is pointed toward us (the observers) where our default is that roughly 1\% of systems have enough angular momentum and the beaming fraction is also 1\%. To constrain these parameters, we compare predictions from our models to the current set of GRB observations.  

We choose a range of theoretical models representing the range of current uncertainties described in Sections~\ref{sec:progenitor},\ref{sec:obs}, including the IMF changes, star-formation history, and other factors related to the theoretical models of GRB formation, such as the binary fraction and stellar wind in the progenitor.  A summary of these models and their parameters is provided in Table \ref{tab:highz_GRB_models}.

\begin{table*}
\begin{center}
\begin{tabular}{|c|c|c|c|c|c|c|}
\hline\hline              
& Model & $f_{\mathrm{IMF}}$ & $z_0$ & $f_{\mathrm{bin}}$ & $f_{\mathrm{wind}}$ & SFR adopted \\
\hline
1 & Mz2.5IMF0.1Bin0.0Wind30 & 0.1 & 2.5 & 0.0 & 30 & \cite{2014ARAA..52..415M} \\
2 & Mz5IMF0.4Bin0.0Wind45 & 0.4 & 5.0 & 0.0 & 45 & \cite{2014ARAA..52..415M} \\
3 & Mz5IMF0.4Bin0.0Wind9000 & 0.4 & 5.0 & 0.0 & 9000 & \cite{2014ARAA..52..415M} \\
4 & Mz5IMF0.0Bin0.1Wind45 & 0.0 & 5.0 & 0.1 & 45 & \cite{2014ARAA..52..415M} \\
5 & Mz5IMF0.0Bin0.1Wind9000 & 0.0 & 5.0 & 0.1 & 9000 & \cite{2014ARAA..52..415M} \\
6 & Mz2.5IMF0.1Bin0.0Wind30 & 0.1 & 2.5 & 0.0 & 30 & \cite{Madau17} \\
7 & Mz2.5IMF0.1Bin0.0Wind30 & 0.1 & 2.5 & 0.0 & 30 & \cite{2006ApJ...651..142H} \\
\hline\hline
\end{tabular}
\caption{Summary table of the theoretical models that are adopted for BAT calibration and GRB prediction for the Gamow Explorer.}
\label{tab:highz_GRB_models}
%\end{indented}
\end{center}
\end{table*}

As we have discussed in section~\ref{sec:prog}, at high metallicity, mass loss from winds reduces the number of systems that collapse to form black holes and, ultimately, the number of GRBs.  In our models, the $f_{\rm wind}$ parameter reduces which massive stars continue to form black holes at high metallicities.  Matching the BAT data to our predicted GRB rate at lower redshift can constrain this parameter.  With a few values of $f_{\rm wind}$, we can use the full BAT dataset to constrain our remaining two parameters:  $f_{\rm GRB}$ and $f_{L_{\rm peak}}$.

$f_{L_{\rm peak}}$ changes the fiducial luminosity $L_{\rm fiducial}$, which is the benchmark luminosity in the luminosity function. The luminosity function shown in Fig.~\ref{fig:grblumdist} has $L_{\rm fiducial} = 10^{52} \ \rm erg \ s^{-1}$ corresponding to an $f_{L_{\rm peak}}=1.0$. In our calibration, we allow $f_{L_{\rm peak}}$ to vary from 0.1-10 to find the $L_{\rm fiducial}$ (between $10^{51}$ and $10^{53}$\,erg) that produces the best-fit results with BAT observations.

For the BAT observed GRBs, we use the GRB sample from the Swift Gamma-Ray Burst Host Galaxy Legacy Survey (``SHOALS'') \citep{Perley16}, which provides the most complete GRB redshift sample from a systematic host-galaxy observations and thus is less affected by biases in redshift measurements. This sample consists of 119 GRBs detected by {\it Swift}/BAT from 2005 to 2012, with a redshift distribution peaking at $z \sim 2$ and the highest redshift at $z \sim 6$. To avoid further uncertainty from the low sample number at higher redshift, we only compare the theoretical models with the redshift distribution in $0<z<4$ from the SHOALS sample. 

To estimate the intrinsic GRB rate for each theoretical model, we multiply the number of GRBs per stellar mass from each model to the star formation rate described in Sect.~\ref{sec:SFR}.  Using the intrinsic GRB rate and the corresponding luminosity function for each theoretical model, we estimate the expected {\it Swift}/BAT GRB detection rate. For GRB spectral shape, we adopt a Band function with $E_{\rm peak}$, $\alpha$, and $\beta$ following distributions described in \citet{Ghirlanda15,GHirlanda21}.
%, with $\sigma = 0.2$ and mean = $-1.0$ and $-2.3$, respectively. 
Note that the $E_{\rm peak}$ from \citet{GHirlanda21} follows the Amati $E_{\rm peak}-L$ relationship, implying an harder spectrum for a more luminous burst.

\begin{figure}[!ht]
%\vspace{-5pt}
\begin{center}
\includegraphics[width=0.55\textwidth]{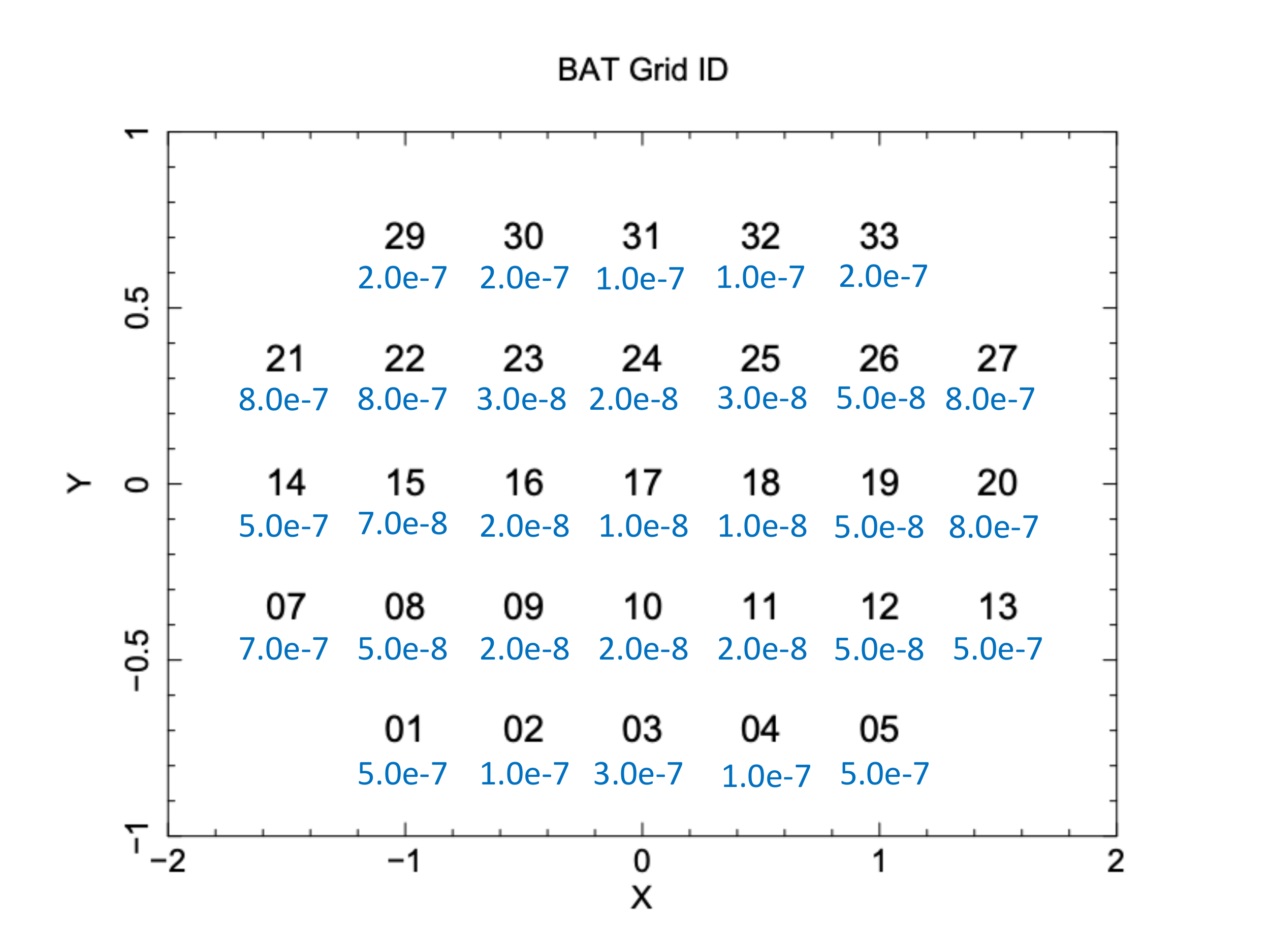}
\end{center}
\includegraphics[width=0.50\textwidth]{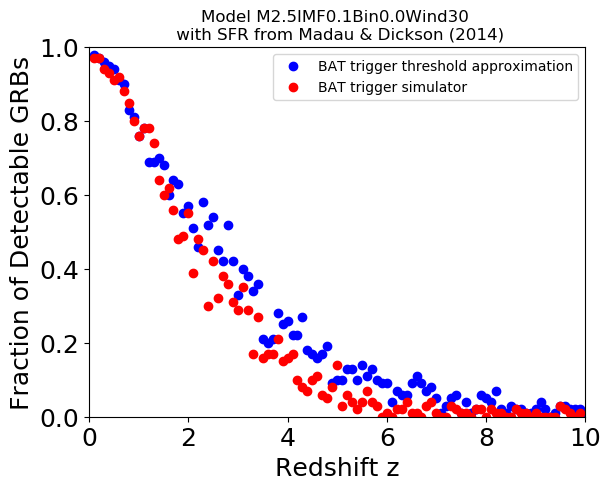}
\caption{{\it Upper Panel:} Approximated flux thresholds (in unit of $\rm erg \ s^{-1} \ cm^{-2}$) for BAT detection criterion for different locations on the BAT detector plane (i.e. the Grid ID, as defined in \citet{Lien14}), which is correlated with a source incident angle.
{\it Lower Panel:} Comparison of the fraction of BAT-detected GRBs between simulations using the approximated flux thresholds and BAT trigger simulator.}
\label{fig:BAT_gridID}
\end{figure}

For each theoretical model, we simulate 14900 GRBs from $0.1 <= z < 15.0$ (i.e., 100 bursts in each redshift bin of 0.1). The corresponding fluxes in BAT energy range, $15-150$ KeV are calculated based on the redshift, luminosity, and spectral shape of each burst.  A simulated GRB would be considered detected by BAT if it passes the detection thresholds. We adopt the BAT detection thresholds presented in \citet{Lien14}, which performed a detailed study of BAT sensitivity of GRBs using the BAT trigger simulator that mimics the BAT onboard trigger algorithm. In order to speed up the simulation to explore a wider parameter range, we adopt an approximated flux detection thresholds for each burst incidental angle, as shown in Fig. 8 in \citet{Lien14}. The upper panel of Fig.~ \ref{fig:BAT_gridID} shows the detection flux limits we adopted based on Fig. 8 in \citet{Lien14}. We have cross-checked this set approximated flux thresholds by running a few of our models with the BAT trigger simulator. As shown in the lower panel of Fig.~\ref{fig:BAT_gridID}, GRBs detected by this set of approximated flux thresholds are in great consistency with GRBs detected by the trigger simulator. For all our simulations, GRB incident angles are randomly assigned to each burst, and the number of active detector in BAT is set to 25000, which is around the medium value throughout the BAT mission time.

Each simulated sample are compared with the real BAT detected GRBs to quantity the goodness of fits. Specifically, we aim to quantify two parameters for each theoretical model: (1) The fiducial luminosity $L_{\rm fiducial}$ using $\chi^2$ fit to the SHOALS GRB redshift sample \citep{Perley16}, and (2) The fraction of stars that forms GRBs ($f_{\rm GRB}$) by adjusting this fraction to match the total BAT detection rate of 90 GRBs per 0.8 year (i.e., 112.5 GRBs/yr). The number 0.8 is the average fraction of survey time that BAT spends on searching for GRBs \citep{Lien16}. 
%Since we assume a canonical beaming factor of $f_{\rm beam} = 0.01$, the value of $f_{\rm disk}$ is estimated as $f_{\rm disk} = f_{\rm GRB}/f_{\rm beam}$.

\begin{figure}[!ht]
%\vspace{-5pt}
\begin{center}
\includegraphics[width=0.45\textwidth]{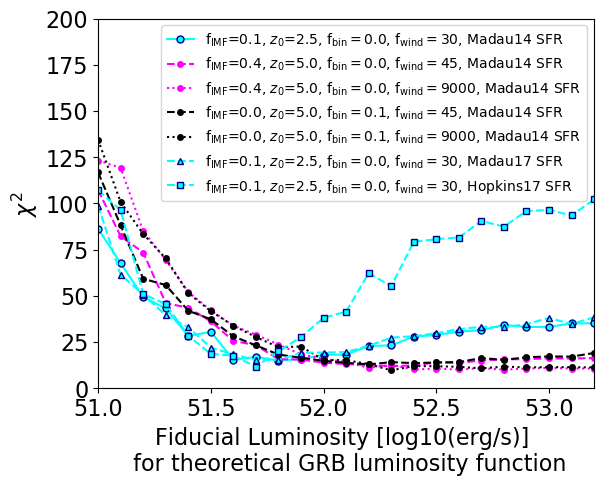}
\end{center}
%\vspace{-18pt}
\caption{
$\chi^2$ for different fiducial luminosity for each theoretical model. 
}
\label{fig:Chi2_fiducial_lum}
\end{figure}

\begin{figure}[!ht]
%\vspace{-5pt}
\begin{center}
\includegraphics[width=0.45\textwidth]{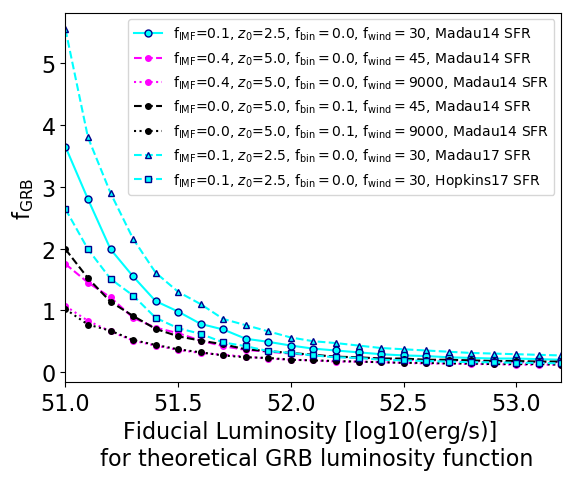}
\end{center}
%\vspace{-18pt}
\caption{
The value of $f_{\rm GRB}$ (the fraction of stars that have sufficient angular momentum to form GRBs and beamed toward us) for different fiducial luminosity for each theoretical model.
}
\label{fig:star2BH_fiducial_lum}
\end{figure}

Figure \ref{fig:Chi2_fiducial_lum} and Figure \ref{fig:star2BH_fiducial_lum} summarize the $\chi^2$ and $f_{\rm GRB}$ values for each model, respectively. 
%These values are also presented in Tables \ref{tab:Chi2_table1} to \ref{tab:fdisk_table2}. 
For most of our models, the best fit to current data lies in a fiducial luminosity ranges from $10^{51.5}\,{\rm erg \, s^{-1}}$ to $10^{52.5}\,{\rm erg \, s^{-1}}$. 
$f_{\rm GRB}$ decreases as the fiducial luminosity increases. This is because if we assume a brighter intrinsic GRB sample, we only need a lower number of intrinsic GRBs to produce the same number of BAT detections. 

\begin{figure}[!ht]
%\vspace{-5pt}
\begin{center}
\includegraphics[width=0.45\textwidth]{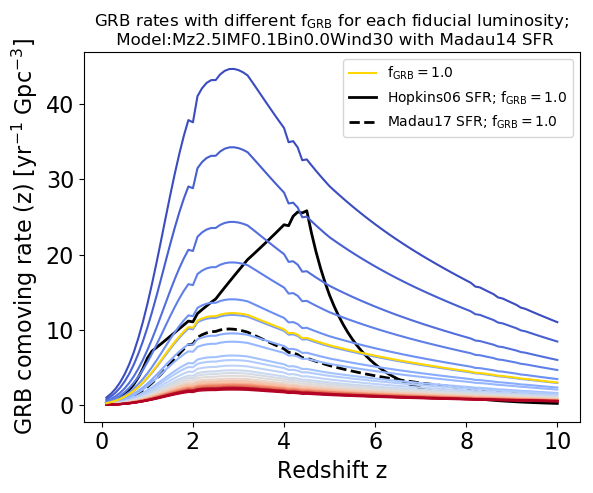}
\includegraphics[width=0.45\textwidth]{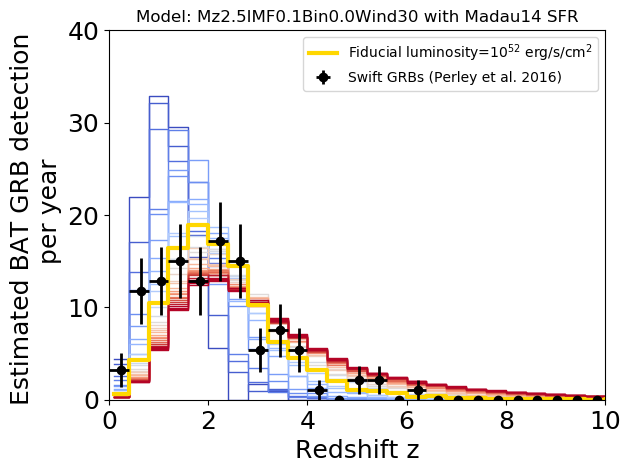}
\end{center}
%\vspace{-18pt}
\caption{
{\it Upper Panel:} GRB comoving rates that corresponds to different $f_{\rm GRB}$ from calibration of BAT data with a range of fiducial luminosity from $10^{51} \ \rm erg \ s^{-1}$ (bluer color) to $10^{54} \ \rm erg \ s^{-1}$ (redder color).
{\it Bottom Panel:} Comparison of the redshift distribution of real BAT detected GRBs and the estimated BAT detections based on theoretical model Mz2.5IMF0.1Bin0.0Wind30 over a range of fiducial luminosity.
}
\label{fig:GRB_model_rate}
\end{figure}

Figure \ref{fig:GRB_model_rate} shows an example of the comoving cosmic GRB rate (upper panel) and the corresponding predicted BAT GRB detection rate for the model Mz2.5IMF0.1Bin0.0Wind30, with a range of fiducial luminosity from $10^{51} \ \rm erg \ s^{-1}$ to $10^{54} \ \rm erg \ s^{-1}$. The upper panel shows that the intrinsic comoving GRB rate required to produce the observed BAT detection decreases as the fiducial luminosity increases, which corresponds to the decrease of $f_{\rm GRB}$ as a function of fiducial luminosity shown in Figure \ref{fig:star2BH_fiducial_lum}. The lower panel presents the comparison of predicted redshift distribution for BAT with the observed redshift distribution from the SHOALS sample. For this model, GRB samples with fiducial luminosities around $\sim 10^{51.8} \ \rm erg \ s^{-1}$ matches the best with the observed redshift distribution.

In addition to the main calibration discussed above, we perform some comparison with the high-redshift GRBs in the SHOALS sample to make sure that the models we included here are in general consistent with the Swift high-redshift GRB detections. In the SHOALS sample, there are 4 GRBs with redshift $z > 5$. In addition, there is one GRB without a redshift measurement and three GRBs that have redshifts consistent with $z > 5$ when including the redshift uncertainty. Therefore, there could be $\sim 4$ to $8$ GRBs with $z > 5$ in the SHOALS sample. Normalizing the SHOALS sample of 119 GRBs to the BAT GRB detections of 112.5/yr, this corresponds to $\sim 3.8$ to $7.6$ GRBs with $z > 5$ per year. We find that for all of our models, there are some fiducial luminosities that produce $3.8$ to $7.6$ GRBs with $z > 5$ per year. For example, a fiducial luminosity in the range of $10^{52.1} \ \rm erg \ s^{-1}$ to $10^{52.2} \ \rm erg \ s^{-1}$ for Model Mz2.5IMF0.1Bin0.0Wind30 gives a BAT high-z GRB numbers that are consistent with this range. For Model Mz5IMF0.4Bin0.0Wind45, a fiducial luminosity greater than $10^{52.7} \ \rm erg \ s^{-1}$ provides high-z GRB rates that is consistent with this range. In general, GRB samples with finducial luminosities below $10^{52.0} \ \rm erg \ s^{-1}$ produces fewer than $3.8$ BAT-detected $z > 5$ GRBs per year. Also, except for model Mz5IMF0.0Bin0.1Wind45 and Mz5IMF0.0Bin0.1Wind9000, GRB samples with finducial luminosities above $10^{53.0} \ \rm erg \ s^{-1}$ produces more than $7.6$ BAT-detected $z > 5$ GRBs per year.

\subsection{Potential contribution of dim GRBs}
\label{sec:dim_GRBs}
Our standard models do not include the potentially large fraction of dim GRBs.  At very low redshift, these dim GRBs can dominate the observed GRB sample.  We have augmented are standard model by extending this dim portion of the GRBs.   The luminosity function for dim GRBs is quite uncertain from both the theoretical and observational point of view.  It is difficult to constrain the nature of these dim brusts.  Dim GRBs could be produced for a wide range of reasons (e.g. off-angle jets or failed jets) and, with our lack of understanding of jet physics, it is difficult to theoretically determine both the rate and properties of these dim GRBs.  Observational constraints are also difficult because these dim GRBs are difficult to constrain because their flux can be mostly below the detection thresholds of the available observatories. To have a crude estimation of the dim GRB contribution in the BAT sample, we explore a dimmer luminosity function by lowering the fiducial luminosity to $f_{\rm fiducial} = 10^{50} \ \rm erg \ s^{-1}$. for dim GRBs.

\begin{figure}[!ht]
%\vspace{-5pt}
\begin{center}
\includegraphics[width=0.45\textwidth]{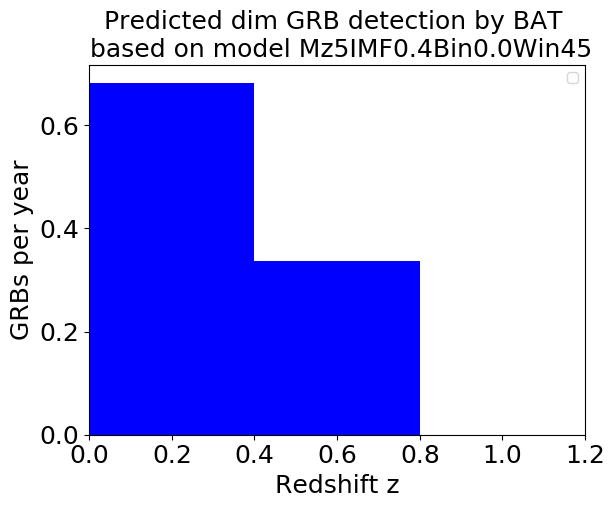}
\end{center}
%\vspace{-18pt}
\caption{
Prediction of dim GRBs in the BAT sample using theoretical model Mz5IMF0.4Bin0.0Win45 and a luminosity function with a lower fiducial luminosity of $10^{50} \ \rm erg \ s^{-1}$. The star-formation rate adopted here is from \cite{2014ARAA..52..415M}.
}
\label{fig:dimGRBs}
\end{figure}

Figure \ref{fig:dimGRBs} shows the expected detections of dim GRBs in the BAT sample based on these two luminosity functions. Based on our current assumptions, results show that these dim GRBs should consist of no more than a few detections per year in the BAT sample (i.e., $\lesssim 1\%$). Dim GRBs are difficult to detect, but should also produce hypernovae.  This places some limits on the relative fraction of dim GRBs to normal GRBs at low redshift.  With these constraints, we do not expect too many dim GRBs in the observed GRB sample.

\section{Studying Constraints from High Redshift}
\label{sec:highz}

\subsection{High Redshift GRBs}

%Figures Needed:  
%1) Prediction for GRB rates above redshift 4 (comparison with current data
%2) Predictions for Gamow

Using the calibrated values of $L_{\rm fiducial}$ and $f_{\rm GRB}$, we estimate detection rates of high-redshift GRBs from a next-generation detector.  For this study, we use the characteristics of the newly proposed Gamow Explorer, a proposed multi-wavelength space telescope aiming for exploring the early universe universe with GRBs.  The instrumental design provided by the Gamow team includes the effective area as a function of energy, the background count rate of $1.632 \times 10^{-5} \rm \ count \ s^{-1} \ arcmin^{-2}$, and the PSF area of $721 \rm \ arcmin^2$.  The Gamow photon counts for each burst $C_{\rm GRB}$ is estimated using the Gamow effective area and the burst energy spectrum.
The total background count during a burst duration is estimated as $C_{\rm bkg} = 1.632 \times 721 \times T_{\rm 90}$, where $T_{90}$ is randomly selected from the BAT-detected GRBs. A burst would be considered detected if the source count $C_{\rm GRB}$ is larger than the photon count that corresponds to a probability of $< 10^{-10}$ from Poisson fluctuation of the total background count $C_{\rm bkg}$. Moreover, a minimum source count of 5 is required, in order to take into account of additional systematic noise. 

% \begin{enumerate}
%     \item Mz5IMF0.0Bin0.1Wind00, the model that produces the lowest GRB rate at high redshift. 
%     \item Mz5IMF0.4Bin0.0Wind00, the model that produces the highest GRB rate at high redshift.
%     \item Mz5IMF0.4Bin0.0Wind90, the model that produces the best-fit result with BAT-detected GRBs.
%     \item Mz5IMF0.4Bin0.0Wind90.sfr5.5, the best-fit model when adjusting high-redshift slope of the star-formation rate.
% \end{enumerate}

\begin{figure*}[!ht]
%\vspace{-5pt}
\begin{center}
\includegraphics[width=1.0\textwidth]{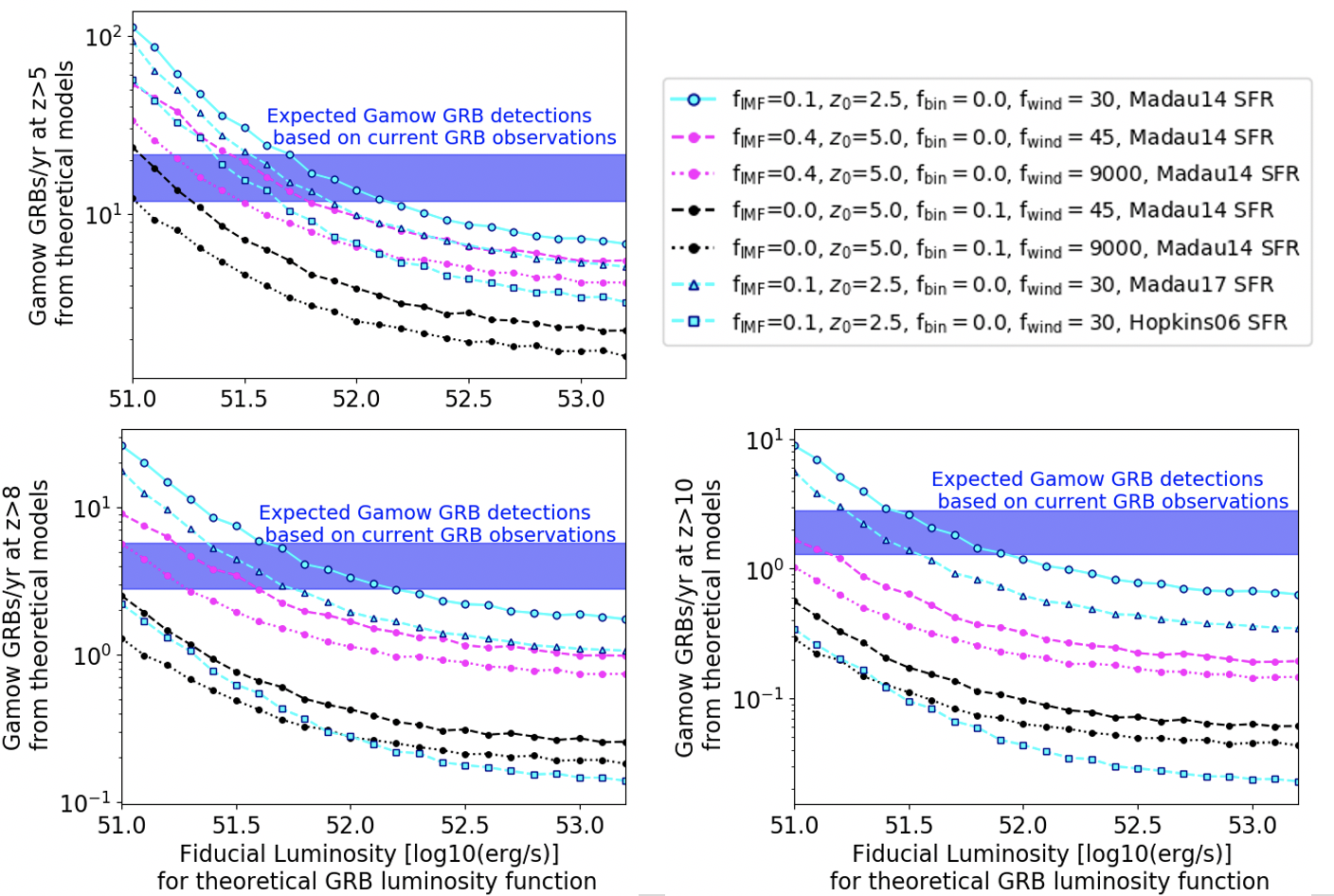}
\end{center}
%\vspace{-18pt}
\caption{
Gamow high-z GRB detection rate for different theoretical models.
}
\label{fig:Gamow_GRB_rate}
\end{figure*}

Figure \ref{fig:Gamow_GRB_rate} shows the estimated Gamow GRB detection rate for these models, and Table \ref{tab:Gamow_GRBs} summarizes the accumulated number of GRB detections for different redshifts.  The differences between these different models shows the true strength of high-redshift observations.  For example, for a fiducial luminosity of $10^{51.5}\,{\rm erg}$, the detection rate above a redshift of 5 ranges from 4-40.  Above a redshift of 8, this rate ranges from a few tenths to ten.  Even a null detection by Gamow will constrain our understanding of our models and the evolution of the IMF.

\begin{figure}[!ht]
%\vspace{-5pt}
\begin{center}
\includegraphics[width=0.45\textwidth]{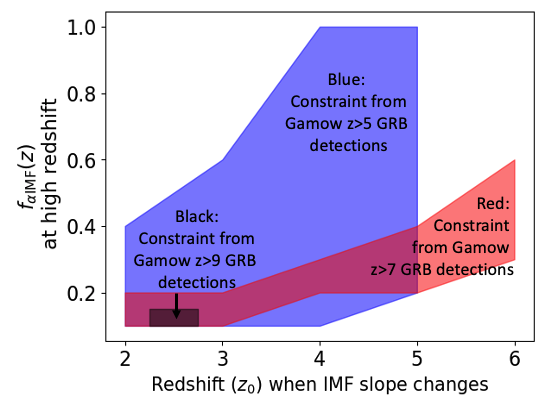}
\end{center}
%\vspace{-18pt}
\caption{
Estimation of the constraints on parameters related to the IMF evolution. The x and y axis show the $z_0$ and $f_{\rm IMF}$ parameters in Eq.~\ref{eq:imf}, respectively. 
}
\label{fig:Gamow_IMF_constraint}
\end{figure}

Figure ~\ref{fig:Gamow_IMF_constraint} shows an example of the expected constraints on parameters related to IMF evolution that could be placed by the Gamow high-z GRB survey.  We assume values for the evolution of the IMF based the current best-estimates of the redshift evolution~\cite{2016A&A...587A..40P}.  A good fit to that redshift evolution finds redshift parameters of $z_0=2.5$ and $f_{\rm IMF}=0.15$ from Eq.~\ref{eq:imf}.  With the current limited high-redshift GRB observations, it is difficult to truly measure this number.  To determine how the added statistics from a mission like Gamow can constrain these IMF evolution parameters, we use a suite of simulations with gradual changes of $f_{\rm IMF}$ from 0.0 to 0.4 and $z_0$ from 2.25 to 1.5. For these additional simulations, all other parameters are fixed to the same values. That is, we use the star-formation rate from \cite{Madau17}, a fiducial luminosity of $10^{51.5} \rm \ erg \ s^{-1} \ cm^2$, binary fraction $f_{\rm bin} = 0.01$, and wind parameter $f_{\rm wind} = 45$. For each of these models, we estimate the Gamow GRB detections at different redshifts and compare that with the Expected Gamow GRB detection rates based on current GRB observations (i.e., the blue regions in Fig.~\ref{fig:Gamow_GRB_rate}). The $f_{\rm IMF}$ and $z_0$ values that give a GRB detection rate within blue region are included in the constraint area shown in Fig.~\ref{fig:Gamow_IMF_constraint}.

\begin{table*}[t]
\begin{center}
\begin{tabular}{|c|c|c|c|c|c|c|c|c|}
\hline\hline              
Model  & SFR adopted & $z>0$ & $z>5$ & $z>6$ & $z>7$ & $z>8$ & $z>9$ & $z>10$ \\
\hline\hline
Mz2.5IMF0.1Bin0.0Wind30 & \cite{2014ARAA..52..415M} & 176.88 & 30.59 & 19.08 & 12.00 & 7.48 & 4.57 & 2.61 \\
\hline
Mz5IMF0.4Bin0.0Win45 & \cite{2014ARAA..52..415M} & 142.93 & 19.77 & 13.33 & 7.30 & 3.46 & 1.50 & 0.64 \\
\hline
Mz5IMF0.4Bin0.0Wind9000 & \cite{2014ARAA..52..415M} & 108.06 & 11.51 & 7.74 & 4.21 & 1.95 & 0.85 & 0.36 \\
\hline
Mz5IMF0.0Bin0.1Win45 & \cite{2014ARAA..52..415M} & 120.51 & 7.14 & 3.28 & 1.59 & 0.77 & 0.37 & 0.17  \\
\hline
Mz5IMF0.0Bin0.1Win9000 & \cite{2014ARAA..52..415M} & 103.74 & 4.59 & 2.06 & 1.00 & 0.49 & 0.24 & 0.11 \\
\hline
Mz2.5IMF0.1Bin0.0Wind30 & \cite{Madau17} & 171.31 & 22.49 & 12.93 & 7.55 & 4.45 & 2.53 & 1.39 \\
\hline
Mz2.5IMF0.1Bin0.0Wind30 & \cite{2006ApJ...651..142H} & 179.09 & 15.42 & 4.73 & 1.67 & 0.62 & 0.24 & 0.09 \\
\hline\hline
\end{tabular}
\caption{Number of predicted GRB detections per year for Gamow for each model.}
\label{tab:Gamow_GRBs}
%\end{indented}
\end{center}
\end{table*}

\subsection{Other High-Redshift Transients:  Supernovae Pair-Instability Supernovae and Hypernovae}

\begin{figure}[!ht]
\plotone{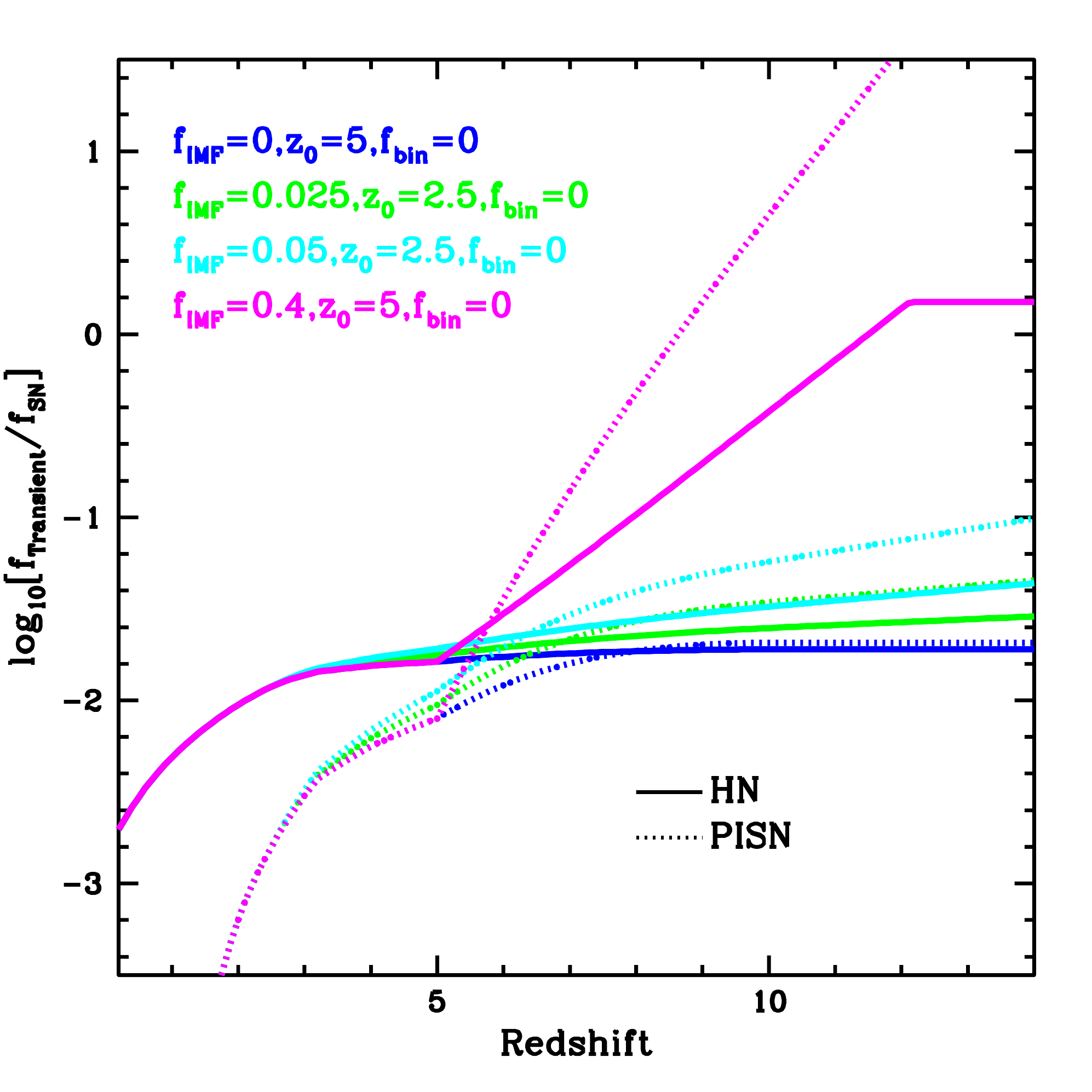}
\caption{Hypernova and Pair-Instability rate with respect to the supernova rate as a function of redshift for 4 of our evolutionary models.  The rate of both of these transients will increase at low redshift as mass loss from winds diminishes.  Because these transients are produced by the most massive stars, if the IMF flattens at high redshift, the rate of these transients increases dramatically and can exceed the supernova rate.} % caption
\label{fig:fpisn}
\end{figure}

In addition to predicting the evolution (both the rate and properties) of GRBs, our models predict the evolution with redshift (due to mass-loss and initial mass function evolution) of other transients produced by massive stars, e.g. pair-instability supernovae and hypernovae.  Figure~\ref{fig:fpisn} shows the rate of pair-instability supernovae and hypernovae with respect to the supernova rate as a function of redshift.  As with gamma-ray bursts, both of these transients are reduced at high metallicity because of mass-loss from metallicity-driven winds.  At low redshifts, there is an increase in the rates of these systems because of the change in winds.  But at high redshifts, the rate can increase even further as the IMF flattens.  In our most extreme case, the pair-instability rate ultimately becomes more than 10 times higher than the supernova rate.  In this section, we discuss this impact and the potential for a broader range of observational constraints on this redshift evolution.  These transients are among the brightest astrophysical outbursts, allowing them to be observed to large distances~\citep{2014ApJ...781..106W,2015arXiv150308147W}.  It is possible that the James Webb Space Telescope~\citep{2006SSRv..123..485G} will be able to observe pair-instability supernovae out to redshift 10~\citep{2014ApJ...781..106W}.  

\begin{figure}[!ht]
\plotone{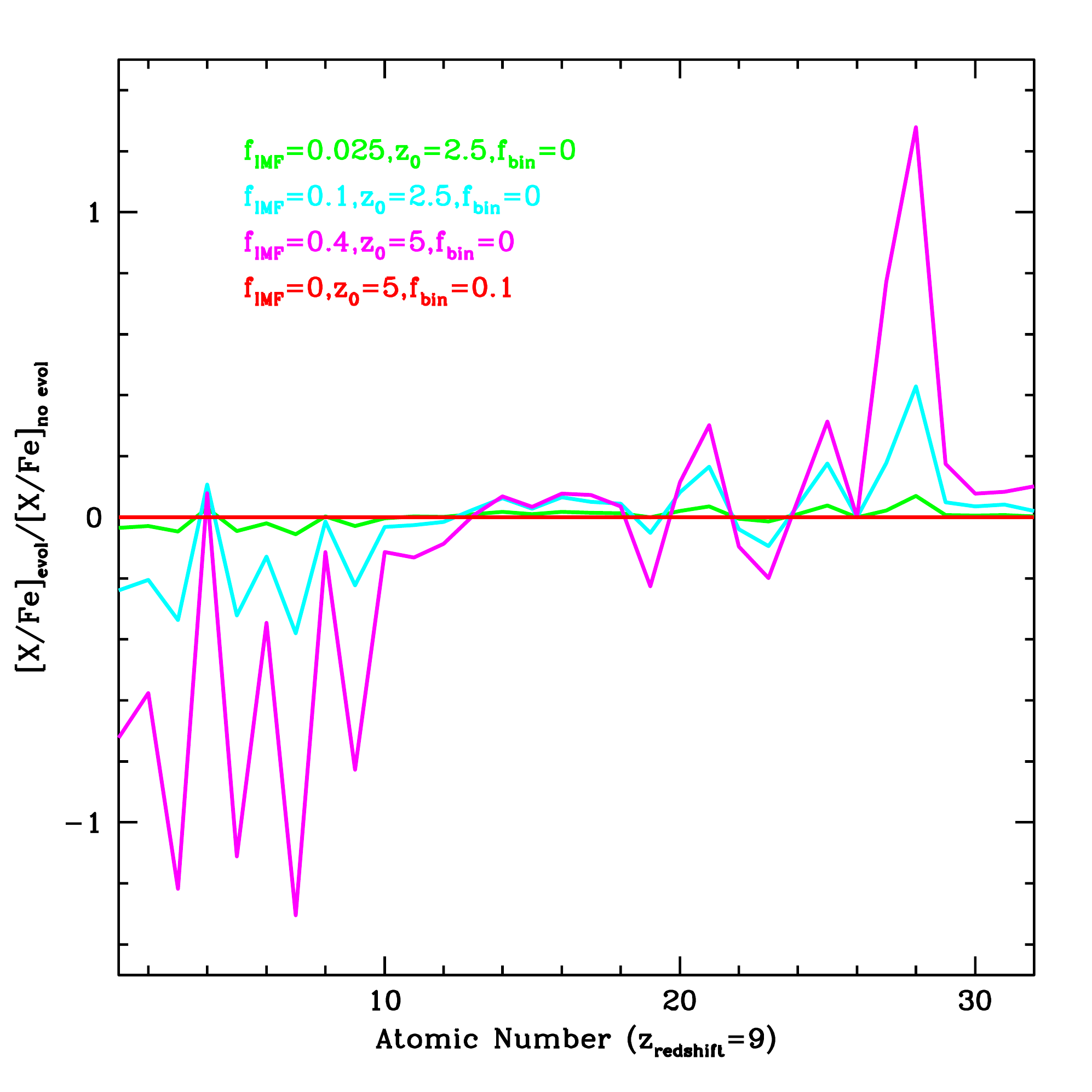}
\caption{Ratio of nucleosynthetic yields at z=9 from our IMF evolving models with respect to a model with no IMF evolution (normalized by the iron production) for atomic numbers ranging from hydrogen to iron peak elements.  As the IMF flattens, a larger fraction of hypernovae and pair-instability supernovae are formed, producing a larger fraction of iron-peak elements.} 
\label{fig:yields}
\end{figure}

Hypernovae and pair-instability supernovae produce different nucleosynthetic yields than normal supernovae and, if the IMF flattens, these unique nucleosynthetic signatures can be compared to observations (e.g. in metal poor stars) to constrain the flattening the IMF at high redshift.  A number of studies have calculate the yields of hypernovae and pair-instability supernovae~\citep{2001ApJ...555..880N,2002ApJ...565..385U,2002ApJ...567..532H,2014A&A...566A.146K,2014ApJ...792...44C}.  Using the yields from \cite{2002ApJ...565..385U}, we can calculate the nucleosynthetic yields as a function of redshift for our different models.  Figure~\ref{fig:yields} shows the yields (abundance divided by iron production:  [X/Fe]) for 4 different IMF evolving models divided by the flat [X/Fe] ratios.  Per star, more iron-peak elements are produced for these models with flattened IMFs.

Other studies have attempted or identified approaches to use metal-poor stars to search for evidence of hypernovae and pair-instability supernova explosions~\cite[e.g.][]{2002ApJ...565..385U,2004ApJ...612..602T,2012RAA....12.1637R}.  In some cases, scientists have argued that the yields of these stars matched hypernovae with little indication for evidence of signatures from pair-instability supernovae~\citep{2002ApJ...565..385U,2004ApJ...612..602T,2018ApJ...857..111T}.  The fact that, thusfar, no clear signature of pair-instability supernovae exists either suggests that we do not know the exact range of masses for pair-instability supernovae or that the IMF does not flatten as much as suggested in our most extreme model.  But much more work must be done to move from finding evidence of these extreme explosions to probing the IMF at high redshifts from these nucleosynthesis observations.

Another probe of the nuclear yields at high-redshift is to identify high-redshift gas (e.g. through active galactic nuclei or gamma-ray bursts) and, through absorption line spectroscopy, probe the nucleosynthetic yields.  Missions like Gamow are designed to obtain these absorption-line spectra to constrain these yields. 

\begin{figure}[!ht]
\plotone{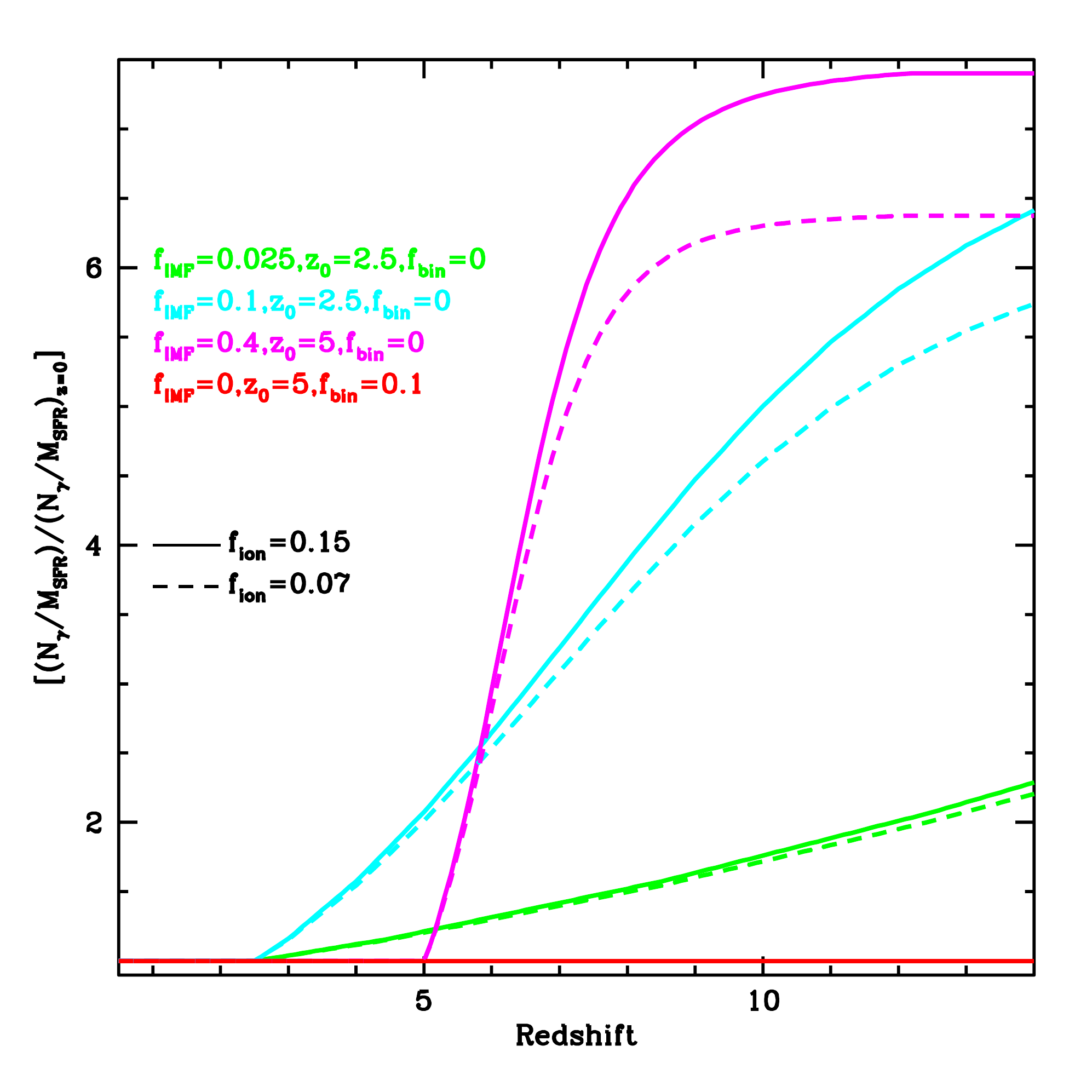}
\caption{Number of Ionizing photons for our different evolving models divided by a model assuming no evolution.  As the IMF flattens out and more massive stars are produced, the fraction of massive stars increases and the number of ionizing photons also increases.} % caption
\label{fig:phion}
\end{figure}

An additional diagnostic of an evolving initial mass function is the number of ionizing photons produced by stars.  Stellar model predictions of the number of ionizing photons produced during the lifetime of massive stars depends on rotation and metallicity, but these differences are actually less than differences produced by different stellar codes~\citep{topping15}.  However, with all models, there is a clear trend where the number of ionizing photons produced increases with stellar mass.  To study the effect of our evolving initial mass functions, we fit the models in \cite{topping15}: 
\begin{equation}
Q_0 = Q_{\rm base} [1 + f_{\rm ion}(M_{\rm star}-20)]
\end{equation}
where $Q_0$ is the life-time integrated H-ionizing photons for a star of mass ($M_{\rm star}$), $Q_{\rm base}$ is roughly $10^{53}$ photons and $f_{\rm ion}$ is determined by fitting to the data in \cite{topping15} ($f_{\rm ion} = 0.07-0.15$).  Figure~\ref{fig:phion} shows the ratio of this number of photons with redshift for 4 of our evolutionary models.

An increase in ionizing photon production from an evolving initial mass function could relieve tension in reionization models where large escape fractions of ionizing photons from galaxies ($>10\%$) are required at high redshift to complete reionization by $z\approx 5.5-6$. Observations of galaxies at $z<4$, where escape fractions of ionizing photons can be measured, predominantly report low escape fractions ($<10\%$) \citep{siana10,sandberg15,rutkowski17,grazian17,pahl21,saxena21}, though recently higher escape fractions have been measured in some galaxies \citep{steidel18,fletcher19,MC21,izotov21,saxena21}. Higher Lyman continuum photon production efficiencies in galaxies at high redshift are favored in \citet{finkelstein19} where they consider scenarios for reionization with empirically-supported low galaxy ionizing photon escape fractions.
 
On the other hand, if Lyman continuum photons are too numerous at high redshift, then reionization will complete sooner than observed quasar sightlines suggest and predicted opacities in the IGM may be too low. This may put constraints on the more extreme models shown in Figure~\ref{fig:phion}, barring a lower escape fraction. However, because the uncertainties in reionization, such as the escape fraction and the emissivity of ionizing photons, are degenerate the incidence of GRBs at high redshift may help constrain each of these properties.

\begin{figure}[!ht]
\plotone{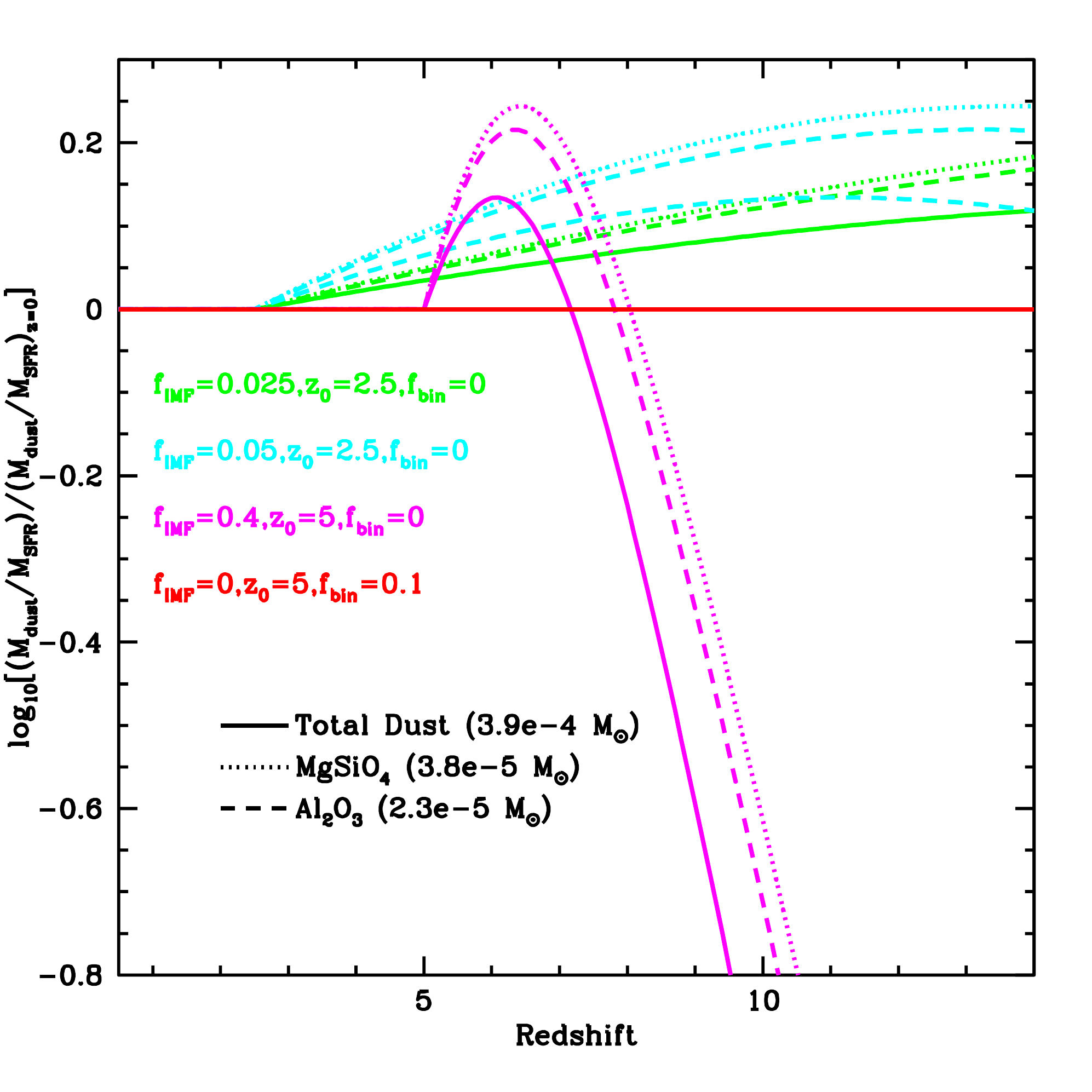}
\caption{Dust as a function of redshift for 4 of our evolutionary models using the dust production from \cite{brooker21}.  Hypernovae produce more dust than normal supernovae so there is an initial rise in dust production with redshift.  We assume no dust production from pair-instability supernovae.  As the flattening of the IMF moves many stars into the pair-instability regime, the dust production decreases.} % caption
\label{fig:dustvz}
\end{figure}

Finally, energetic explosions such as hypernovae may have an enhanced dust production rate at high redshift~\citep{brooker21}.  There is some evidence of this increased dust production at low metallicities~\citep{2020AA...641A.168N}.  Figure~\ref{fig:dustvz} shows the total dust (plus two specific dust grains) production using the models from \cite{brooker21} and our IMF/metallicity evolution.  Initially, the dust production increases before decreasing as we make increasingly massive stars.  This production does not include any contribution from pair-instability supernovae which may be a major contributor to high-redshift dust production~\cite{2004MNRAS.351.1379S}.  If they produce large amounts of dust, these results would have to be revised at high redshift.

Coupled with a larger sample of GRB observations probing the same redshift space, this broad suite of observations will be able to set firm limits on the evolution of the IMF at high redshift.

\section{Summary}

This paper outlines the conditions that lead to an evolution of the rate and properties of GRBs with redshift under the BHAD paradigm where collapsars are the primary progenitor behind long-duration GRBs.  In our models, the primary factors in determining the rate include an evolving IMF and differences in binary interactions.  By measuring the GRB rate at high redshift, astronomers can probe both the IMF and the star formation history.

For current standard star formation histories, the number of GRBs that should be observed above a redshift of 8 by a mission like Gamow is less than one if the is no IMF evolution but could be above ten for extensive IMF evolution.  There is a limit to which a flattening of the IMF can increase the rate of GRBs.  Standard GRBs are not easily produced by very massive stars (the black hole is too massive to produce strong jets).  If the IMF flattens sufficiently such that most stars are formed with masses above 100\,M$_\odot$ where the stars tend to preferentially form pair-instability supernovae or very massive stars that form weak GRBs, the GRB rate will decrease. The variation in the rate comparing different star formation histories is also an order of magnitude.  The Gamow probe is ideally suited to making the measurements needed to constrain the GRB rate and star formation history.

For our standard collapsar GRB progenitor, the rate is primarily set by the IMF and the SFH and it is difficult to distinguish these two effects from the rate alone.  However, the properties of GRBs and other transients do depend on the IMF and, by measuring these properties (e.g. observations of the transients, nucleosynthetic yields, dust production), we can probe not only the IMF, but also the star formation history at high redshift.

In this paper, we also discussed many of the uncertainties and the caveats in our models.  First and foremost of these is that we don't truly know the exact progenitor for gamma-ray bursts nor do we understand entirely the mechanism producing GRB jets.  It is indeed possible that the range of solutions predicted in this paper are superseded by the arrival of a new progenitor scenario.  If observations push for trends beyond what is predicted in our models, we may instead discover key insight into the engines and/or progenitors of these cosmic explosions.



\acknowledgements
The work by CLF, JLJ, and PRUS was supported by the US Department of Energy through the Los Alamos National Laboratory. Los Alamos National Laboratory is operated by Triad National Security, LLC, for the National Nuclear Security Administration of U.S.\ Department of Energy (Contract No.\ 89233218CNA000001).  The {\it Gamow} mission proposal to NASA was an essential motivation for this work. The {\it Gamow} PI Nicholas E. White and the entire  science team are thanked for their support and essential discussions. 

\bibliography{refs}{}

\begin{thebibliography}{}
\expandafter\ifx\csname natexlab\endcsname\relax\def\natexlab#1{#1}\fi
\providecommand{\url}[1]{\href{#1}{#1}}
\providecommand{\dodoi}[1]{doi:~\href{http://doi.org/#1}{\nolinkurl{#1}}}
\providecommand{\doeprint}[1]{\href{http://ascl.net/#1}{\nolinkurl{http://ascl.net/#1}}}
\providecommand{\doarXiv}[1]{\href{https://arxiv.org/abs/#1}{\nolinkurl{https://arxiv.org/abs/#1}}}

\bibitem[{{Arnett}(1980)}]{1980ApJ...237..541A}
{Arnett}, W.~D. 1980, \apj, 237, 541, \dodoi{10.1086/157898}

\bibitem[{{Atteia}(1997)}]{1997A&A...328L..21A}
{Atteia}, J.~L. 1997, \aap, 328, L21.
\newblock \doarXiv{astro-ph/9708238}

\bibitem[{{Barkat} {et~al.}(1967){Barkat}, {Rakavy}, \&
  {Sack}}]{1967PhRvL..18..379B}
{Barkat}, Z., {Rakavy}, G., \& {Sack}, N. 1967, \prl, 18, 379,
  \dodoi{10.1103/PhysRevLett.18.379}

\bibitem[{{Belczynski} {et~al.}(2010){Belczynski}, {Holz}, {Fryer}, {Berger},
  {Hartmann}, \& {O'Shea}}]{2010ApJ...708..117B}
{Belczynski}, K., {Holz}, D.~E., {Fryer}, C.~L., {et~al.} 2010, \apj, 708, 117,
  \dodoi{10.1088/0004-637X/708/1/117}

\bibitem[{{Belczynski} {et~al.}(2020){Belczynski}, {Klencki}, {Fields},
  {Olejak}, {Berti}, {Meynet}, {Fryer}, {Holz}, {O'Shaughnessy}, {Brown},
  {Bulik}, {Leung}, {Nomoto}, {Madau}, {Hirschi}, {Kaiser}, {Jones}, {Mondal},
  {Chruslinska}, {Drozda}, {Gerosa}, {Doctor}, {Giersz}, {Ekstrom}, {Georgy},
  {Askar}, {Baibhav}, {Wysocki}, {Natan}, {Farr}, {Wiktorowicz}, {Coleman
  Miller}, {Farr}, \& {Lasota}}]{2020AA...636A.104B}
{Belczynski}, K., {Klencki}, J., {Fields}, C.~E., {et~al.} 2020, \aap, 636,
  A104, \dodoi{10.1051/0004-6361/201936528}

\bibitem[{{Bloom} {et~al.}(1999){Bloom}, {Sigurdsson}, \&
  {Pols}}]{1999MNRAS.305..763B}
{Bloom}, J.~S., {Sigurdsson}, S., \& {Pols}, O.~R. 1999, \mnras, 305, 763,
  \dodoi{10.1046/j.1365-8711.1999.02437.x}

\bibitem[{{Bouwens} {et~al.}(2020){Bouwens}, {Gonz{\'a}lez-L{\'o}pez},
  {Aravena}, {Decarli}, {Novak}, {Stefanon}, {Walter}, {Boogaard}, {Carilli},
  {Dudzevi{\v{c}}i{\={u}}t{\.{e}}}, {Smail}, {Daddi}, {da Cunha}, {Ivison},
  {Nanayakkara}, {Cortes}, {Cox}, {Inami}, {Oesch}, {Popping}, {Riechers}, {van
  der Werf}, {Weiss}, {Fudamoto}, \& {Wagg}}]{2020ApJ...902..112B}
{Bouwens}, R., {Gonz{\'a}lez-L{\'o}pez}, J., {Aravena}, M., {et~al.} 2020,
  \apj, 902, 112, \dodoi{10.3847/1538-4357/abb830}

\bibitem[{{Bromm}(2013)}]{2013RPPh...76k2901B}
{Bromm}, V. 2013, Reports on Progress in Physics, 76, 112901,
  \dodoi{10.1088/0034-4885/76/11/112901}

\bibitem[{{Bromm} {et~al.}(2001){Bromm}, {Ferrara}, {Coppi}, \&
  {Larson}}]{2001MNRAS.328..969B}
{Bromm}, V., {Ferrara}, A., {Coppi}, P.~S., \& {Larson}, R.~B. 2001, \mnras,
  328, 969, \dodoi{10.1046/j.1365-8711.2001.04915.x}

\bibitem[{{Brooker} {et~al.}(2021){Brooker}, {Stangl}, {Mauney}, \&
  {Fryer}}]{brooker21}
{Brooker}, E.~S., {Stangl}, S.~M., {Mauney}, C.~M., \& {Fryer}, C.~L. 2021,
  arXiv e-prints, arXiv:2103.12781.
\newblock \doarXiv{2103.12781}

\bibitem[{{Calder{\'o}n} {et~al.}(2020){Calder{\'o}n}, {Cuadra}, {Schartmann},
  {Burkert}, {Prieto}, \& {Russell}}]{2020MNRAS.493..447C}
{Calder{\'o}n}, D., {Cuadra}, J., {Schartmann}, M., {et~al.} 2020, \mnras, 493,
  447, \dodoi{10.1093/mnras/staa090}

\bibitem[{{Chabrier}(2003)}]{2003PASP..115..763C}
{Chabrier}, G. 2003, \pasp, 115, 763, \dodoi{10.1086/376392}

\bibitem[{{Chen} {et~al.}(2014){Chen}, {Heger}, {Woosley}, {Almgren}, \&
  {Whalen}}]{2014ApJ...792...44C}
{Chen}, K.-J., {Heger}, A., {Woosley}, S., {Almgren}, A., \& {Whalen}, D.~J.
  2014, \apj, 792, 44, \dodoi{10.1088/0004-637X/792/1/44}

\bibitem[{{Chon} {et~al.}(2021){Chon}, {Omukai}, \&
  {Schneider}}]{2021arXiv210304997C}
{Chon}, S., {Omukai}, K., \& {Schneider}, R. 2021, arXiv e-prints,
  arXiv:2103.04997.
\newblock \doarXiv{2103.04997}

\bibitem[{{Costa} {et~al.}(2020){Costa}, {Bressan}, {Mapelli}, {Marigo},
  {Iorio}, \& {Spera}}]{2020arXiv201002242C}
{Costa}, G., {Bressan}, A., {Mapelli}, M., {et~al.} 2020, arXiv e-prints,
  arXiv:2010.02242.
\newblock \doarXiv{2010.02242}

\bibitem[{{De La Rosa} {et~al.}(2017){De La Rosa}, {Roming}, \&
  {Fryer}}]{2017ApJ...850..133D}
{De La Rosa}, J., {Roming}, P., \& {Fryer}, C. 2017, \apj, 850, 133,
  \dodoi{10.3847/1538-4357/aa93ee}

\bibitem[{{Dominik} {et~al.}(2013){Dominik}, {Belczynski}, {Fryer}, {Holz},
  {Berti}, {Bulik}, {Mand el}, \& {O'Shaughnessy}}]{2013ApJ...779...72D}
{Dominik}, M., {Belczynski}, K., {Fryer}, C., {et~al.} 2013, \apj, 779, 72,
  \dodoi{10.1088/0004-637X/779/1/72}

\bibitem[{{Farmer} {et~al.}(2020){Farmer}, {Renzo}, {de Mink}, {Fishbach}, \&
  {Justham}}]{2020ApJ...902L..36F}
{Farmer}, R., {Renzo}, M., {de Mink}, S.~E., {Fishbach}, M., \& {Justham}, S.
  2020, \apjl, 902, L36, \dodoi{10.3847/2041-8213/abbadd}

\bibitem[{{Finkelstein} {et~al.}(2019){Finkelstein}, {D'Aloisio},
  {Paardekooper}, {Ryan}, {Behroozi}, {Finlator}, {Livermore}, {Upton
  Sanderbeck}, {Dalla Vecchia}, \& {Khochfar}}]{finkelstein19}
{Finkelstein}, S.~L., {D'Aloisio}, A., {Paardekooper}, J.-P., {et~al.} 2019,
  \apj, 879, 36, \dodoi{10.3847/1538-4357/ab1ea8}

\bibitem[{{Fletcher} {et~al.}(2019){Fletcher}, {Tang}, {Robertson}, {Nakajima},
  {Ellis}, {Stark}, \& {Inoue}}]{fletcher19}
{Fletcher}, T.~J., {Tang}, M., {Robertson}, B.~E., {et~al.} 2019, \apj, 878,
  87, \dodoi{10.3847/1538-4357/ab2045}

\bibitem[{{Fontana} {et~al.}(2006){Fontana}, {Salimbeni}, {Grazian},
  {Giallongo}, {Pentericci}, {Nonino}, {Fontanot}, {Menci}, {Monaco},
  {Cristiani}, {Vanzella}, {de Santis}, \& {Gallozzi}}]{2006AA...459..745F}
{Fontana}, A., {Salimbeni}, S., {Grazian}, A., {et~al.} 2006, \aap, 459, 745,
  \dodoi{10.1051/0004-6361:20065475}

\bibitem[{{Frey} {et~al.}(2013){Frey}, {Fryer}, \&
  {Young}}]{2013ApJ...773L...7F}
{Frey}, L.~H., {Fryer}, C.~L., \& {Young}, P.~A. 2013, \apjl, 773, L7,
  \dodoi{10.1088/2041-8205/773/1/L7}

\bibitem[{{Fryer}(1999)}]{1999ApJ...522..413F}
{Fryer}, C.~L. 1999, \apj, 522, 413, \dodoi{10.1086/307647}

\bibitem[{{Fryer} {et~al.}(2012){Fryer}, {Belczynski}, {Wiktorowicz},
  {Dominik}, {Kalogera}, \& {Holz}}]{2012ApJ...749...91F}
{Fryer}, C.~L., {Belczynski}, K., {Wiktorowicz}, G., {et~al.} 2012, \apj, 749,
  91, \dodoi{10.1088/0004-637X/749/1/91}

\bibitem[{{Fryer} {et~al.}(2019){Fryer}, {Lloyd-Ronning}, {Wollaeger},
  {Wiggins}, {Miller}, {Dolence}, {Ryan}, \& {Fields}}]{2019EPJA...55..132F}
{Fryer}, C.~L., {Lloyd-Ronning}, N., {Wollaeger}, R., {et~al.} 2019, European
  Physical Journal A, 55, 132, \dodoi{10.1140/epja/i2019-12818-y}

\bibitem[{{Fryer} {et~al.}(2014){Fryer}, {Rueda}, \&
  {Ruffini}}]{2014ApJ...793L..36F}
{Fryer}, C.~L., {Rueda}, J.~A., \& {Ruffini}, R. 2014, \apjl, 793, L36,
  \dodoi{10.1088/2041-8205/793/2/L36}

\bibitem[{{Fryer} {et~al.}(1999){Fryer}, {Woosley}, \&
  {Hartmann}}]{1999ApJ...526..152F}
{Fryer}, C.~L., {Woosley}, S.~E., \& {Hartmann}, D.~H. 1999, \apj, 526, 152,
  \dodoi{10.1086/307992}

\bibitem[{{Fryer} {et~al.}(2001){Fryer}, {Woosley}, \&
  {Heger}}]{2001ApJ...550..372F}
{Fryer}, C.~L., {Woosley}, S.~E., \& {Heger}, A. 2001, \apj, 550, 372,
  \dodoi{10.1086/319719}

\bibitem[{{Fryer} {et~al.}(2007){Fryer}, {Mazzali}, {Prochaska}, {Cappellaro},
  {Panaitescu}, {Berger}, {van Putten}, {van den Heuvel}, {Young},
  {Hungerford}, {Rockefeller}, {Yoon}, {Podsiadlowski}, {Nomoto}, {Chevalier},
  {Schmidt}, \& {Kulkarni}}]{2007PASP..119.1211F}
{Fryer}, C.~L., {Mazzali}, P.~A., {Prochaska}, J., {et~al.} 2007, \pasp, 119,
  1211, \dodoi{10.1086/523768}

\bibitem[{{Gal-Yam} \& {Maoz}(2004)}]{2004MNRAS.347..942G}
{Gal-Yam}, A., \& {Maoz}, D. 2004, \mnras, 347, 942,
  \dodoi{10.1111/j.1365-2966.2004.07237.x}

\bibitem[{{Gardner} {et~al.}(2006){Gardner}, {Mather}, {Clampin}, {Doyon},
  {Greenhouse}, {Hammel}, {Hutchings}, {Jakobsen}, {Lilly}, {Long}, {Lunine},
  {McCaughrean}, {Mountain}, {Nella}, {Rieke}, {Rieke}, {Rix}, {Smith},
  {Sonneborn}, {Stiavelli}, {Stockman}, {Windhorst}, \&
  {Wright}}]{2006SSRv..123..485G}
{Gardner}, J.~P., {Mather}, J.~C., {Clampin}, M., {et~al.} 2006, \ssr, 123,
  485, \dodoi{10.1007/s11214-006-8315-7}

\bibitem[{{Ghirlanda} {et~al.}(2004){Ghirlanda}, {Ghisellini}, {Lazzati}, \&
  {Firmani}}]{Ghirlanda2004}
{Ghirlanda}, G., {Ghisellini}, G., {Lazzati}, D., \& {Firmani}, C. 2004, \apjl,
  613, L13, \dodoi{10.1086/424915}

\bibitem[{{Ghirlanda} {et~al.}(2007){Ghirlanda}, {Nava}, {Ghisellini}, \&
  {Firmani}}]{2007A&A...466..127G}
{Ghirlanda}, G., {Nava}, L., {Ghisellini}, G., \& {Firmani}, C. 2007, \aap,
  466, 127, \dodoi{10.1051/0004-6361:20077119}

\bibitem[{{Ghirlanda} {et~al.}(2015){Ghirlanda}, {Salvaterra}, {Ghisellini},
  {Mereghetti}, {Tagliaferri}, {Campana}, {Osborne}, {O'Brien}, {Tanvir},
  {Willingale}, {Amati}, {Basa}, {Bernardini}, {Burlon}, {Covino}, {D'Avanzo},
  {Frontera}, {G{\"o}tz}, {Melandri}, {Nava}, {Piro}, \&
  {Vergani}}]{Ghirlanda15}
{Ghirlanda}, G., {Salvaterra}, R., {Ghisellini}, G., {et~al.} 2015, \mnras,
  448, 2514, \dodoi{10.1093/mnras/stv183}

\bibitem[{{Ghirlanda} {et~al.}(2021){Ghirlanda}, {Salvaterra}, {Toffano},
  {Ronchini}, {Guidorzi}, {Oganesyan}, {Ascenzi}, {Bernardini}, {Camisasca},
  {Mereghetti}, {Nava}, {Ravasio}, {Branchesi}, {Castro-Tirado}, {Amati},
  {Blain}, {Bozzo}, {O'Brien}, {G{\"o}tz}, {Le Floch}, {Osborne}, {Rosati},
  {Stratta}, {Tanvir}, {Bogomazov}, {D'Avanzo}, {Hafizi}, {Mandhai},
  {Melandri}, {Peer}, {Topinka}, {Vergani}, \& {Zane}}]{GHirlanda21}
{Ghirlanda}, G., {Salvaterra}, R., {Toffano}, M., {et~al.} 2021, Experimental
  Astronomy, \dodoi{10.1007/s10686-021-09763-3}

\bibitem[{{Grazian} {et~al.}(2017){Grazian}, {Giallongo}, {Paris}, {Boutsia},
  {Dickinson}, {Santini}, {Windhorst}, {Jansen}, {Cohen}, {Ashcraft},
  {Scarlata}, {Rutkowski}, {Vanzella}, {Cusano}, {Cristiani}, {Giavalisco},
  {Ferguson}, {Koekemoer}, {Grogin}, {Castellano}, {Fiore}, {Fontana},
  {Marchi}, {Pedichini}, {Pentericci}, {Amor{\'\i}n}, {Barro}, {Bonchi},
  {Bongiorno}, {Faber}, {Fumana}, {Galametz}, {Guaita}, {Kocevski}, {Merlin},
  {Nonino}, {O'Connell}, {Pilo}, {Ryan}, {Sani}, {Speziali}, {Testa}, {Weiner},
  \& {Yan}}]{grazian17}
{Grazian}, A., {Giallongo}, E., {Paris}, D., {et~al.} 2017, \aap, 602, A18,
  \dodoi{10.1051/0004-6361/201730447}

\bibitem[{Guetta \& Valle(2007)}]{Guetta_2007}
Guetta, D., \& Valle, M.~D. 2007, The Astrophysical Journal, 657, L73,
  \dodoi{10.1086/511417}

\bibitem[{{Hamuy} \& {Pinto}(2002)}]{2002ApJ...566L..63H}
{Hamuy}, M., \& {Pinto}, P.~A. 2002, \apjl, 566, L63, \dodoi{10.1086/339676}

\bibitem[{{Heger} {et~al.}(2003){Heger}, {Fryer}, {Woosley}, {Langer}, \&
  {Hartmann}}]{2003ApJ...591..288H}
{Heger}, A., {Fryer}, C.~L., {Woosley}, S.~E., {Langer}, N., \& {Hartmann},
  D.~H. 2003, \apj, 591, 288, \dodoi{10.1086/375341}

\bibitem[{{Heger} \& {Woosley}(2002)}]{2002ApJ...567..532H}
{Heger}, A., \& {Woosley}, S.~E. 2002, \apj, 567, 532, \dodoi{10.1086/338487}

\bibitem[{{Hopkins} \& {Beacom}(2006)}]{2006ApJ...651..142H}
{Hopkins}, A.~M., \& {Beacom}, J.~F. 2006, \apj, 651, 142,
  \dodoi{10.1086/506610}

\bibitem[{{Iwamoto} {et~al.}(1998){Iwamoto}, {Mazzali}, {Nomoto}, {Umeda},
  {Nakamura}, {Patat}, {Danziger}, {Young}, {Suzuki}, {Shigeyama},
  {Augusteijn}, {Doublier}, {Gonzalez}, {Boehnhardt}, {Brewer}, {Hainaut},
  {Lidman}, {Leibundgut}, {Cappellaro}, {Turatto}, {Galama}, {Vreeswijk},
  {Kouveliotou}, {van Paradijs}, {Pian}, {Palazzi}, \&
  {Frontera}}]{1998Natur.395..672I}
{Iwamoto}, K., {Mazzali}, P.~A., {Nomoto}, K., {et~al.} 1998, \nat, 395, 672,
  \dodoi{10.1038/27155}

\bibitem[{{Izotov} {et~al.}(2021){Izotov}, {Worseck}, {Schaerer}, {Guseva},
  {Chisholm}, {Thuan}, {Fricke}, \& {Verhamme}}]{izotov21}
{Izotov}, Y.~I., {Worseck}, G., {Schaerer}, D., {et~al.} 2021, \mnras, 503,
  1734, \dodoi{10.1093/mnras/stab612}

\bibitem[{{Kennicutt} \& {Evans}(2012)}]{2012ARAA..50..531K}
{Kennicutt}, R.~C., \& {Evans}, N.~J. 2012, \araa, 50, 531,
  \dodoi{10.1146/annurev-astro-081811-125610}

\bibitem[{{Kozyreva} {et~al.}(2014){Kozyreva}, {Yoon}, \&
  {Langer}}]{2014A&A...566A.146K}
{Kozyreva}, A., {Yoon}, S.~C., \& {Langer}, N. 2014, \aap, 566, A146,
  \dodoi{10.1051/0004-6361/201423641}

\bibitem[{{Kroupa}(2001)}]{2001MNRAS.322..231K}
{Kroupa}, P. 2001, \mnras, 322, 231, \dodoi{10.1046/j.1365-8711.2001.04022.x}

\bibitem[{{Kudritzki}(2002)}]{2002ApJ...577..389K}
{Kudritzki}, R.~P. 2002, \apj, 577, 389, \dodoi{10.1086/342178}

\bibitem[{{Lamb} \& {Reichart}(2000)}]{2000ApJ...536....1L}
{Lamb}, D.~Q., \& {Reichart}, D.~E. 2000, \apj, 536, 1, \dodoi{10.1086/308918}

\bibitem[{{Lazar} \& {Bromm}(2021)}]{2021arXiv211011956L}
{Lazar}, A., \& {Bromm}, V. 2021, arXiv e-prints, arXiv:2110.11956.
\newblock \doarXiv{2110.11956}

\bibitem[{{Lien} {et~al.}(2014){Lien}, {Sakamoto}, {Gehrels}, {Palmer},
  {Barthelmy}, {Graziani}, \& {Cannizzo}}]{Lien14}
{Lien}, A., {Sakamoto}, T., {Gehrels}, N., {et~al.} 2014, \apj, 783, 24,
  \dodoi{10.1088/0004-637X/783/1/24}

\bibitem[{{Lien} {et~al.}(2016){Lien}, {Sakamoto}, {Barthelmy}, {Baumgartner},
  {Cannizzo}, {Chen}, {Collins}, {Cummings}, {Gehrels}, {Krimm}, {Markwardt},
  {Palmer}, {Stamatikos}, {Troja}, \& {Ukwatta}}]{Lien16}
{Lien}, A., {Sakamoto}, T., {Barthelmy}, S.~D., {et~al.} 2016, \apj, 829, 7,
  \dodoi{10.3847/0004-637X/829/1/7}

\bibitem[{{Livio} \& {Mazzali}(2018)}]{2018PhR...736....1L}
{Livio}, M., \& {Mazzali}, P. 2018, \physrep, 736, 1,
  \dodoi{10.1016/j.physrep.2018.02.002}

\bibitem[{{Lloyd-Ronning} {et~al.}(2002){Lloyd-Ronning}, {Fryer}, \&
  {Ramirez-Ruiz}}]{2002ApJ...574..554L}
{Lloyd-Ronning}, N.~M., {Fryer}, C.~L., \& {Ramirez-Ruiz}, E. 2002, \apj, 574,
  554, \dodoi{10.1086/341059}

\bibitem[{{Lloyd-Ronning} {et~al.}(2020){Lloyd-Ronning}, {Johnson}, \&
  {Aykutalp}}]{2020MNRAS.498.5041L}
{Lloyd-Ronning}, N.~M., {Johnson}, J.~L., \& {Aykutalp}, A. 2020, \mnras, 498,
  5041, \dodoi{10.1093/mnras/staa2787}

\bibitem[{{Madau} \& {Dickinson}(2014)}]{2014ARAA..52..415M}
{Madau}, P., \& {Dickinson}, M. 2014, \araa, 52, 415,
  \dodoi{10.1146/annurev-astro-081811-125615}

\bibitem[{{Madau} {et~al.}(1996){Madau}, {Ferguson}, {Dickinson}, {Giavalisco},
  {Steidel}, \& {Fruchter}}]{1996MNRAS.283.1388M}
{Madau}, P., {Ferguson}, H.~C., {Dickinson}, M.~E., {et~al.} 1996, \mnras, 283,
  1388, \dodoi{10.1093/mnras/283.4.1388}

\bibitem[{{Madau} \& {Fragos}(2017)}]{Madau17}
{Madau}, P., \& {Fragos}, T. 2017, \apj, 840, 39,
  \dodoi{10.3847/1538-4357/aa6af9}

\bibitem[{{Marques-Chaves} {et~al.}(2021){Marques-Chaves}, {Schaerer},
  {{\'A}lvarez-M{\'a}rquez}, {Colina}, {Dessauges-Zavadsky},
  {P{\'e}rez-Fournon}, {Saldana-Lopez}, \& {Verhamme}}]{MC21}
{Marques-Chaves}, R., {Schaerer}, D., {{\'A}lvarez-M{\'a}rquez}, J., {et~al.}
  2021, \mnras, 507, 524, \dodoi{10.1093/mnras/stab2187}

\bibitem[{{Miller} \& {Scalo}(1979)}]{1979ApJS...41..513M}
{Miller}, G.~E., \& {Scalo}, J.~M. 1979, \apjs, 41, 513, \dodoi{10.1086/190629}

\bibitem[{{Nakamura} {et~al.}(2001){Nakamura}, {Umeda}, {Iwamoto}, {Nomoto},
  {Hashimoto}, {Hix}, \& {Thielemann}}]{2001ApJ...555..880N}
{Nakamura}, T., {Umeda}, H., {Iwamoto}, K., {et~al.} 2001, \apj, 555, 880,
  \dodoi{10.1086/321495}

\bibitem[{{Nanni} {et~al.}(2020){Nanni}, {Burgarella}, {Theul{\'e}},
  {C{\^o}t{\'e}}, \& {Hirashita}}]{2020AA...641A.168N}
{Nanni}, A., {Burgarella}, D., {Theul{\'e}}, P., {C{\^o}t{\'e}}, B., \&
  {Hirashita}, H. 2020, \aap, 641, A168, \dodoi{10.1051/0004-6361/202037833}

\bibitem[{{Nugis} \& {Lamers}(2000)}]{2000AA...360..227N}
{Nugis}, T., \& {Lamers}, H.~J.~G.~L.~M. 2000, \aap, 360, 227

\bibitem[{{O'Connor} \& {Ott}(2011)}]{2011ApJ...730...70O}
{O'Connor}, E., \& {Ott}, C.~D. 2011, \apj, 730, 70,
  \dodoi{10.1088/0004-637X/730/2/70}

\bibitem[{{Pahl} {et~al.}(2021){Pahl}, {Shapley}, {Steidel}, {Chen}, \&
  {Reddy}}]{pahl21}
{Pahl}, A.~J., {Shapley}, A., {Steidel}, C.~C., {Chen}, Y., \& {Reddy}, N.~A.
  2021, \mnras, 505, 2447, \dodoi{10.1093/mnras/stab1374}

\bibitem[{{Pei} {et~al.}(1999){Pei}, {Fall}, \& {Hauser}}]{1999ApJ...522..604P}
{Pei}, Y.~C., {Fall}, S.~M., \& {Hauser}, M.~G. 1999, \apj, 522, 604,
  \dodoi{10.1086/307674}

\bibitem[{{Perley} {et~al.}(2016){Perley}, {Kr{\"u}hler}, {Schulze}, {de Ugarte
  Postigo}, {Hjorth}, {Berger}, {Cenko}, {Chary}, {Cucchiara}, {Ellis}, {Fong},
  {Fynbo}, {Gorosabel}, {Greiner}, {Jakobsson}, {Kim}, {Laskar}, {Levan},
  {Micha{\l}owski}, {Milvang-Jensen}, {Tanvir}, {Th{\"o}ne}, \&
  {Wiersema}}]{Perley16}
{Perley}, D.~A., {Kr{\"u}hler}, T., {Schulze}, S., {et~al.} 2016, \apj, 817, 7,
  \dodoi{10.3847/0004-637X/817/1/7}

\bibitem[{{Perlmutter} {et~al.}(1999){Perlmutter}, {Aldering}, {Goldhaber},
  {Knop}, {Nugent}, {Castro}, {Deustua}, {Fabbro}, {Goobar}, {Groom}, {Hook},
  {Kim}, {Kim}, {Lee}, {Nunes}, {Pain}, {Pennypacker}, {Quimby}, {Lidman},
  {Ellis}, {Irwin}, {McMahon}, {Ruiz-Lapuente}, {Walton}, {Schaefer}, {Boyle},
  {Filippenko}, {Matheson}, {Fruchter}, {Panagia}, {Newberg}, {Couch}, \&
  {Project}}]{1999ApJ...517..565P}
{Perlmutter}, S., {Aldering}, G., {Goldhaber}, G., {et~al.} 1999, \apj, 517,
  565, \dodoi{10.1086/307221}

\bibitem[{{Pescalli} {et~al.}(2015){Pescalli}, {Ghirlanda}, {Salafia},
  {Ghisellini}, {Nappo}, \& {Salvaterra}}]{2015MNRAS.447.1911P}
{Pescalli}, A., {Ghirlanda}, G., {Salafia}, O.~S., {et~al.} 2015, \mnras, 447,
  1911, \dodoi{10.1093/mnras/stu2482}

\bibitem[{{Pescalli} {et~al.}(2016){Pescalli}, {Ghirlanda}, {Salvaterra},
  {Ghisellini}, {Vergani}, {Nappo}, {Salafia}, {Melandri}, {Covino}, \&
  {G{\"o}tz}}]{2016A&A...587A..40P}
{Pescalli}, A., {Ghirlanda}, G., {Salvaterra}, R., {et~al.} 2016, \aap, 587,
  A40, \dodoi{10.1051/0004-6361/201526760}

\bibitem[{{Popham} {et~al.}(1999){Popham}, {Woosley}, \&
  {Fryer}}]{1999ApJ...518..356P}
{Popham}, R., {Woosley}, S.~E., \& {Fryer}, C. 1999, \apj, 518, 356,
  \dodoi{10.1086/307259}

\bibitem[{{Ren} {et~al.}(2012){Ren}, {Christlieb}, \&
  {Zhao}}]{2012RAA....12.1637R}
{Ren}, J., {Christlieb}, N., \& {Zhao}, G. 2012, Research in Astronomy and
  Astrophysics, 12, 1637, \dodoi{10.1088/1674-4527/12/12/005}

\bibitem[{{Riess} {et~al.}(1998){Riess}, {Filippenko}, {Challis},
  {Clocchiatti}, {Diercks}, {Garnavich}, {Gilliland}, {Hogan}, {Jha},
  {Kirshner}, {Leibundgut}, {Phillips}, {Reiss}, {Schmidt}, {Schommer},
  {Smith}, {Spyromilio}, {Stubbs}, {Suntzeff}, \&
  {Tonry}}]{1998AJ....116.1009R}
{Riess}, A.~G., {Filippenko}, A.~V., {Challis}, P., {et~al.} 1998, \aj, 116,
  1009, \dodoi{10.1086/300499}

\bibitem[{{Rosen} \& {Krumholz}(2020)}]{2020AJ....160...78R}
{Rosen}, A.~L., \& {Krumholz}, M.~R. 2020, \aj, 160, 78,
  \dodoi{10.3847/1538-3881/ab9abf}

\bibitem[{{Rutkowski} {et~al.}(2017){Rutkowski}, {Scarlata}, {Henry}, {Hayes},
  {Mehta}, {Hathi}, {Cohen}, {Windhorst}, {Koekemoer}, {Teplitz}, {Haardt}, \&
  {Siana}}]{rutkowski17}
{Rutkowski}, M.~J., {Scarlata}, C., {Henry}, A., {et~al.} 2017, \apjl, 841,
  L27, \dodoi{10.3847/2041-8213/aa733b}

\bibitem[{{Salpeter}(1955)}]{1955ApJ...121..161S}
{Salpeter}, E.~E. 1955, \apj, 121, 161, \dodoi{10.1086/145971}

\bibitem[{{Salvaterra} {et~al.}(2012){Salvaterra}, {Campana}, {Vergani},
  {Covino}, {D'Avanzo}, {Fugazza}, {Ghirlanda}, {Ghisellini}, {Melandri},
  {Nava}, {Sbarufatti}, {Flores}, {Piranomonte}, \&
  {Tagliaferri}}]{2012ApJ...749...68S}
{Salvaterra}, R., {Campana}, S., {Vergani}, S.~D., {et~al.} 2012, \apj, 749,
  68, \dodoi{10.1088/0004-637X/749/1/68}

\bibitem[{{Sandberg} {et~al.}(2015){Sandberg}, {{\"O}stlin}, {Melinder}, {Bik},
  \& {Guaita}}]{sandberg15}
{Sandberg}, A., {{\"O}stlin}, G., {Melinder}, J., {Bik}, A., \& {Guaita}, L.
  2015, \apjl, 814, L10, \dodoi{10.1088/2041-8205/814/1/L10}

\bibitem[{{Saxena} {et~al.}(2021){Saxena}, {Pentericci}, {Ellis}, {Guaita},
  {Calabr{\`o}}, {Schaerer}, {Vanzella}, {Amor{\'\i}n}, {Bolzonella},
  {Castellano}, {Fontanot}, {Hathi}, {Hibon}, {Llerena}, {Mannucci},
  {Saldana-Lopez}, {Talia}, \& {Zamorani}}]{saxena21}
{Saxena}, A., {Pentericci}, L., {Ellis}, R.~S., {et~al.} 2021, arXiv e-prints,
  arXiv:2109.03662.
\newblock \doarXiv{2109.03662}

\bibitem[{{Schneider} {et~al.}(2004){Schneider}, {Ferrara}, \&
  {Salvaterra}}]{2004MNRAS.351.1379S}
{Schneider}, R., {Ferrara}, A., \& {Salvaterra}, R. 2004, \mnras, 351, 1379,
  \dodoi{10.1111/j.1365-2966.2004.07876.x}

\bibitem[{{Siana} {et~al.}(2010){Siana}, {Teplitz}, {Ferguson}, {Brown},
  {Giavalisco}, {Dickinson}, {Chary}, {de Mello}, {Conselice}, {Bridge},
  {Gardner}, {Colbert}, \& {Scarlata}}]{siana10}
{Siana}, B., {Teplitz}, H.~I., {Ferguson}, H.~C., {et~al.} 2010, \apj, 723,
  241, \dodoi{10.1088/0004-637X/723/1/241}

\bibitem[{{Steidel} {et~al.}(2018){Steidel}, {Bogosavljevi{\'c}}, {Shapley},
  {Reddy}, {Rudie}, {Pettini}, {Trainor}, \& {Strom}}]{steidel18}
{Steidel}, C.~C., {Bogosavljevi{\'c}}, M., {Shapley}, A.~E., {et~al.} 2018,
  \apj, 869, 123, \dodoi{10.3847/1538-4357/aaed28}

\bibitem[{{Surman} \& {McLaughlin}(2004)}]{2004ApJ...603..611S}
{Surman}, R., \& {McLaughlin}, G.~C. 2004, \apj, 603, 611,
  \dodoi{10.1086/381672}

\bibitem[{{Takahashi} {et~al.}(2018){Takahashi}, {Yoshida}, \&
  {Umeda}}]{2018ApJ...857..111T}
{Takahashi}, K., {Yoshida}, T., \& {Umeda}, H. 2018, \apj, 857, 111,
  \dodoi{10.3847/1538-4357/aab95f}

\bibitem[{{Topping} \& {Shull}(2015)}]{topping15}
{Topping}, M.~W., \& {Shull}, J.~M. 2015, \apj, 800, 97,
  \dodoi{10.1088/0004-637X/800/2/97}

\bibitem[{{Tremonti} {et~al.}(2004){Tremonti}, {Heckman}, {Kauffmann},
  {Brinchmann}, {Charlot}, {White}, {Seibert}, {Peng}, {Schlegel}, {Uomoto},
  {Fukugita}, \& {Brinkmann}}]{2004ApJ...613..898T}
{Tremonti}, C.~A., {Heckman}, T.~M., {Kauffmann}, G., {et~al.} 2004, \apj, 613,
  898, \dodoi{10.1086/423264}

\bibitem[{{Tumlinson} {et~al.}(2004){Tumlinson}, {Venkatesan}, \&
  {Shull}}]{2004ApJ...612..602T}
{Tumlinson}, J., {Venkatesan}, A., \& {Shull}, J.~M. 2004, \apj, 612, 602,
  \dodoi{10.1086/422571}

\bibitem[{{Umeda} \& {Nomoto}(2002)}]{2002ApJ...565..385U}
{Umeda}, H., \& {Nomoto}, K. 2002, \apj, 565, 385, \dodoi{10.1086/323946}

\bibitem[{{Whalen} {et~al.}(2014){Whalen}, {Smidt}, {Even}, {Woosley}, {Heger},
  {Stiavelli}, \& {Fryer}}]{2014ApJ...781..106W}
{Whalen}, D.~J., {Smidt}, J., {Even}, W., {et~al.} 2014, \apj, 781, 106,
  \dodoi{10.1088/0004-637X/781/2/106}

\bibitem[{{Wiggins} {et~al.}(2015){Wiggins}, {Smidt}, {Whalen}, {Even},
  {Migenes}, \& {Fryer}}]{2015arXiv150308147W}
{Wiggins}, B.~K., {Smidt}, J.~M., {Whalen}, D.~J., {et~al.} 2015, arXiv
  e-prints, arXiv:1503.08147.
\newblock \doarXiv{1503.08147}

\bibitem[{{Woosley} {et~al.}(2002){Woosley}, {Heger}, \&
  {Weaver}}]{2002RvMP...74.1015W}
{Woosley}, S.~E., {Heger}, A., \& {Weaver}, T.~A. 2002, Reviews of Modern
  Physics, 74, 1015, \dodoi{10.1103/RevModPhys.74.1015}

\bibitem[{{Young} \& {Fryer}(2007)}]{2007ApJ...670..584Y}
{Young}, P.~A., \& {Fryer}, C.~L. 2007, \apj, 670, 584, \dodoi{10.1086/521695}

\end{thebibliography}
\bibliographystyle{aasjournal}



\end{document}